\documentclass[iop,tighten]{emulateapj}
\usepackage{natbib}
\usepackage{graphicx}
\usepackage{amssymb}
\usepackage{amsmath}
\usepackage{enumerate}
\usepackage{xcolor}
\usepackage{hyperref}
\hypersetup{
	colorlinks = true,
	linkcolor = black,
	citecolor = blue,
	filecolor= black,
	runcolor= black,
	urlcolor= blue
}

\defcitealias{strom2017}{Strom17}
\defcitealias{steidel2016}{Steidel16}
\defcitealias{pilyugin2012}{Pil12}

\shorttitle{Measuring O/H, N/O, $U$, and Fe/H in high-$z$ galaxies}
\shortauthors{Strom et al.}

\begin{document}

\title{Measuring the Physical Conditions in High-Redshift Star-Forming Galaxies: \\Insights from KBSS-MOSFIRE}
\email{astrom@carnegiescience.edu}
\author{Allison L. Strom\altaffilmark{1,2*}}
\altaffiltext{*}{Carnegie Fellow}
\altaffiltext{1}{Carnegie Observatories, 813 Santa Barbara Street, Pasadena, CA 91101, USA}
\author{Charles C. Steidel\altaffilmark{2}}
\altaffiltext{2}{Cahill Center for Astronomy and Astrophysics, California Institute of Technology, MS 249-17, Pasadena, CA 91125, USA}
\author{Gwen C. Rudie\altaffilmark{1}}
\author{Ryan F. Trainor\altaffilmark{3}}
\altaffiltext{3}{Department of Physics \& Astronomy, Franklin \& Marshall College}
\author{Max Pettini\altaffilmark{4}}
\altaffiltext{4}{Institute of Astronomy, Madingley Road, Cambridge CB3 OHA, UK}
\begin{abstract}
We use photoionization models designed to reconcile the joint rest-UV-optical spectra of high-$z$ star-forming galaxies to self-consistently infer the gas chemistry and nebular ionization and excitation conditions for $\sim150$ galaxies from the Keck Baryonic Structure Survey (KBSS), using only observations of their rest-optical nebular spectra. We find that the majority of $z\sim2-3$ KBSS galaxies are moderately O-rich, with an interquartile range in $12+\log(\textrm{O/H})=8.29-8.56$, and have significantly sub-solar Fe enrichment, with an interquartile range of [Fe/H]$=[-0.79,-0.53]$, contributing additional evidence in favor of super-solar O/Fe in high-$z$ galaxies. Model-inferred ionization parameter and N/O are strongly correlated with common strong-line indices (such as O32 and N2O2), with the latter exhibiting similar behavior to local extragalactic \ion{H}{2} regions. In contrast, diagnostics commonly used for measuring gas-phase O/H (such as N2 and O3N2) show relatively large scatter with the overall amount of oxygen present in the gas and behave differently than observed at $z\sim0$. We provide a new calibration for using R23 to measure O/H in typical high-$z$ galaxies, although it is most useful for relatively O-rich galaxies; combining O32 and R23 does not yield a more effective calibration. Finally, we consider implications for the intrinsic correlations between physical conditions across the galaxy sample and find that N/O varies with O/H in high-$z$ galaxies in a manner almost identical to local \ion{H}{2} regions. However, we do not find a strong anti-correlation between ionization parameter and metallicity (O/H or Fe/H) in high-$z$ galaxies, which is one of the principal bases for using strong-line ratios to infer oxygen abundance.
\end{abstract}
\keywords{cosmology: observations --- galaxies: evolution --- galaxies: high-redshift --- galaxies: ISM --- ISM: abundances --- ISM: \ion{H}{2} regions}
\maketitle

\section{Introduction}

Advancing our understanding of galaxy assembly is one of the key goals of modern astrophysics. However, progress is often difficult, as galaxies form and evolve under the influence of a variety of competing baryonic processes, the effects of which are difficult to disentangle. Gaseous inflows supply the raw material for star formation and the growth of supermassive black holes. The resulting powerful stellar winds, supernova explosions, and feedback from active galactic nuclei (AGN) are all thought to contribute to galaxy-scale outflows, which are relatively uncommon in nearby galaxies, but known to be nearly ubiquitous in the early universe. The details of these processes and their relative importance throughout cosmic time leave imprints on nascent galaxies, resulting in the scaling relations and chemical abundance patterns observed across galaxy populations.

Efforts by several groups over the last few years have extended our understanding of galaxies' physical conditions to the peak of galaxy assembly \citep[$z\sim1-3$; e.g.,][]{madau2014}, using new spectroscopic observations of large numbers of typical galaxies to study their gas and stars in detail \citep[e.g.,][]{masters2014,steidel2014,shapley2015,wisnioski2015,sanders2016,steidel2016,kashino2017,strom2017}. These studies have focused on measurements from galaxies' rest-optical ($3600-7000$\AA) nebular spectra, which can now be observed for large samples of individual objects, owing to sensitive multi-object near-infrared (NIR) spectrographs like the Multi-Object Spectrometer For InfraRed Exploration \citep[MOSFIRE,][]{mclean2012,steidel2014} and the $K$-band Multi-Object Spectrograph \citep[KMOS,][]{sharples2013}. However, despite the headway made in characterizing galaxies during this crucial epoch, significant tension remains regarding how best to infer high-$z$ galaxies' physical conditions---especially chemical abundances---from easily-observable quantities, such as the strong emission lines of hydrogen, oxygen, and nitrogen present in their \ion{H}{2} region spectra.

It is tempting to simply build on the large body of work that has decrypted the spectra of galaxies in the local universe \citep[e.g.,][]{kauffmann2003,brinchmann2008,masters2016}. Such efforts have frequently relied on the sample of $z\sim0$ star-forming galaxies from the Sloan Digital Sky Survey \citep[SDSS,][]{york2000} to explore trends in physical conditions and construct diagnostics for quantities like oxygen abundance (O/H) that can then be applied to observations of other samples, including those at high redshift. Yet it is well-established that star-forming galaxies at $z\sim1-3$ differ from typical $z\sim0$ star-forming galaxies in a number of key ways that make directly transferring this paradigm for understanding galaxies' nebular spectra to the study of the high-$z$ universe problematic: $z\sim2-3$ galaxies have 10~times higher star-formation rates \citep[e.g.,][]{erb2006mass} and gas masses \citep{tacconi2013} at fixed stellar mass, significantly smaller physical sizes \citep{law2012}, and are relatively young (characteristic ages of a few hundred Myr) with rising star-formation histories \citep{reddy2012}. Recent work from surveys like the Keck Baryonic Structure Survey \citep[KBSS,][]{steidel2014,steidel2016,strom2017}, the MOSFIRE Deep Evolution Field survey \citep[MOSDEF,][]{kriek2015,shapley2015}, and the KMOS$^{\textrm{3D}}$ survey \citep{wisnioski2015} have also revealed a number of important differences in terms of the nebular spectra of high-$z$ galaxies, with perhaps the most well-known being the offset in the log([\ion{O}{3}]$\lambda5008$/H$\beta$) vs. log([\ion{N}{2}]$\lambda6585$/H$\alpha$) plane (the so-called N2-BPT diagram, after \citealt{baldwin1981}, but popularized by \citealt{veilleux1987}). This offset has been attributed to a variety of astrophysical differences, including enhanced N/O at fixed O/H in high-$z$ galaxies \citep{masters2014,shapley2015,sanders2016}, higher ionization parameters \citep[e.g.][]{kewley2015,bian2016,kashino2017}, higher electron densities \citep{liu2008,bian2010}, and harder ionizing radiation fields \citep{steidel2014,steidel2016,strom2017}, although the true origin of the offset is likely due to a combination of effects for individual galaxies \citep[see also][]{kojima2017}.

Since strong-line diagnostics operate by relying on the underlying correlations between the quantity of interest (often ``metallicity", or gas-phase O/H) and other astrophysical conditions (e.g., ionization state, ionizing photon distribution) that also influence the observables, it is imperative to consider how these quantities are different in typical high-$z$ galaxies relative to present-day galaxies \textit{and} how they may vary among high-$z$ galaxies. In \citet[][hereafter Steidel16]{steidel2016}, we sought to directly address this issue by combining observations of the rest-UV spectra of high-$z$ galaxies' massive stellar populations with observations of the ionized gas surrounding the same stars, as probed by their rest-UV-optical nebular spectra. We compared a composite rest-UV-optical spectrum of 30 star-forming galaxies from KBSS with stellar population models and photoionization model predictions and showed that only models that simultaneously include iron-poor massive star binaries and moderate oxygen enrichment in the gas can reconcile all of the observational constraints, even accounting for somewhat higher ionization parameters and electron densities than observed in typical $z\sim0$ galaxies. In \citet[][hereafter Strom17]{strom2017}, we found that this abundance pattern---low Fe/H and moderate-to-high O/H---is also necessary to explain the behavior of \textit{individual} $z\simeq2-2.7$ KBSS galaxies and the behavior of the high-$z$ star-forming galaxy locus in multiple 2D line-ratio spaces, including the N2-BPT diagram, the S2-BPT diagram (which trades [\ion{S}{2}]$6718,6732$ for [\ion{N}{2}]$\lambda6585$), and the O32-R23 diagram\footnote{O32=$\log(\textrm{[\ion{O}{3}]}\lambda\lambda4960,5008/\textrm{[\ion{O}{2}]}\lambda\lambda3727,3729$) and R23=$\log[(\textrm{[\ion{O}{3}]}\lambda\lambda4960,5008+\textrm{[\ion{O}{2}]}\lambda\lambda3727,3729)/\textrm{H}\beta]$.} (which is sensitive to changes in ionization and excitation).

In this paper, we expand on the analysis from \citetalias{steidel2016} and \citetalias{strom2017} and present a new method for self-consistently determining gas-phase oxygen abundance (O/H), nitrogen-to-oxygen ratio (N/O), and ionization parameter ($U$) in individual high-$z$ galaxies, utilizing measurements of the nebular emission lines in their rest-optical spectra and photoionization models motivated by observations of the same high-$z$ galaxies. The sample used here is described in Section~\ref{sample_selection}, followed by a description of the photoionization model method in Section~\ref{model_section}. The physical parameters inferred for individual galaxies are presented in Section~\ref{results_section}. The results of this analysis are used to determine new strong-line diagnostics for $U$, N/O, and O/H in Section~\ref{strongline_section}, along with our guidance regarding the best method for determining these quantities using emission line measurements. New constraints on relationships between physical conditions in high-$z$ galaxies' \ion{H}{2} regions (including the N/O-O/H relation) are presented in Section~\ref{correlation_section}. We conclude with a summary of our results in Section~\ref{summary_section}.

Throughout the paper, we adopt the solar metallicity scale from \citet{asplund2009}, with $Z_{\odot} = 0.0142$, 12+log(Fe/H)$_{\odot}=7.50$, $12+\log(\textrm{O/H})$$_{\odot}=8.69$, and log(N/O)$_{\odot} = -0.86$. When necessary, we assume a $\Lambda$CDM cosmology: $H_0=70$\,km s$^{-1}$ Mpc$^{-1}$, $\Omega_{\Lambda}=0.7$, and $\Omega_{\textrm{m}} = 0.3$. Finally, specific spectral features are referred to using their vacuum wavelengths in angstroms.

\section{Sample Description}
\label{sample_selection}

The sample of galaxies we consider for analysis is drawn from KBSS, which is a large, spectroscopic galaxy survey conducted in 15 fields centered on bright quasars and is explicitly designed to study the galaxy-gas correlation during the peak of cosmic star formation \citep{rudie2012,steidel2014}. Extensive imaging and spectroscopic campaigns have been conducted in all of the KBSS fields, including deep broad-band and medium-band optical-IR imaging\footnote{All fields have imaging in $U_n$, $G$, $\mathcal{R}$, $J$, and $K_s$ from the ground, as well as data from \textit{Hubble}/WFC3-IR F140W and \textit{Spitzer}/IRAC at $3.6\mu$m and $4.5\mu$m. 10 fields have at least one deep pointing obtained using \textit{Hubble}/WFC3-IR F160W, and 8 fields have NIR imaging in $J1$, $J2$, $J3$, $H1$, and $H2$ collected using Magellan-FourStar \citep{persson2013}.}, as well as spectroscopy obtained using the Low Resolution Imaging Spectrometer \citep[LRIS,][]{oke1995,steidel2004} at optical wavelengths and MOSFIRE in the NIR bands. The acquisition and reduction of the photometric and spectroscopic data have been described elsewhere \citep[e.g., by][]{steidel2003,reddy2012,steidel2014}, with additional details related to the rest-optical spectroscopic analysis (including the measurement of emission line fluxes and corrections for relative slit-losses between the NIR bands) provided by \citetalias{strom2017}.

\begin{figure}
\centering
\includegraphics{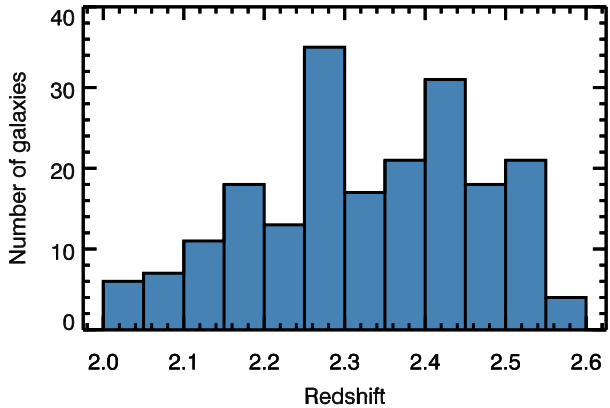}
\caption{The distribution of nebular redshift for the sample of 202 galaxies discussed in this paper, which has $\langle z\rangle = 2.3$.}
\label{sample_zhist}
\end{figure}

For this paper, we selected the subsample of $z\simeq2-2.7$ KBSS galaxies with nebular redshifts measured from MOSFIRE spectra ($\langle z\rangle = 2.3$, Figure~\ref{sample_zhist}) and spectral coverage of the regions near H$\alpha$, H$\beta$, [\ion{O}{3}]$\lambda5008$, and [\ion{N}{2}]$\lambda6585$ (i.e., the lines used in the N2-BPT diagram). Although not required for inclusion in the sample, measurements of or limits on [\ion{O}{2}]$\lambda\lambda3727,3729$, [\ion{O}{3}]$\lambda4364$, [\ion{S}{2}]$\lambda\lambda6718,6732$ and [\ion{Ne}{3}]$\lambda3869$ are incorporated when present. Objects are included regardless of the signal-to-noise ratio (SNR) for a single line measurement, but we restrict the sample to those galaxies with $\textrm{SNR}>5$ measurements of the Balmer decrement (H$\alpha$/H$\beta$), which we use to correct for dust attenuation. This requirement is imposed to ensure a fair comparison with the line flux predictions from photoionization models, which represent the nebular spectrum as it would appear if unattenuated by dust along the line-of-sight. We also exclude galaxies where there is evidence of significant AGN activity \citep[usually in the form of high-ionization rest-UV emission features; see][]{steidel2014,strom2017}, as these galaxies will not be well-matched by photoionization model predictions made using ionizing radiation fields from purely stellar sources. In total, 202~galaxies satisfy these criteria and compose the sample discussed in the remainder of the paper.

\begin{figure}
\centering
\includegraphics{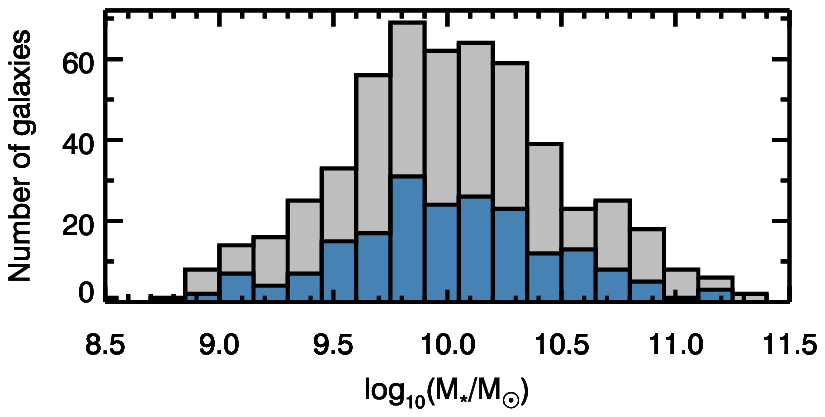}
\includegraphics{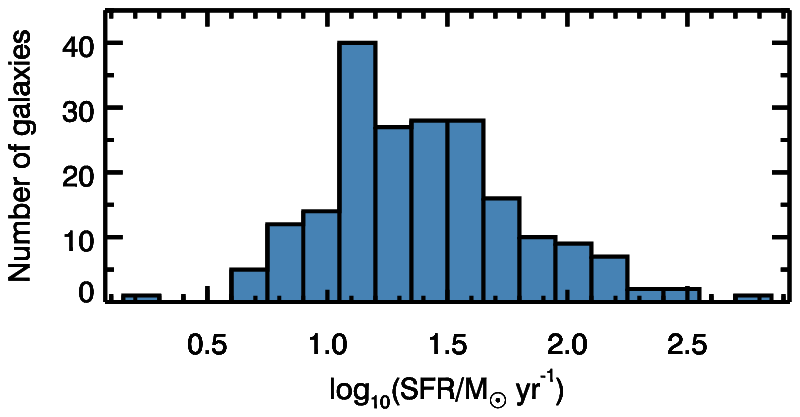}
\includegraphics{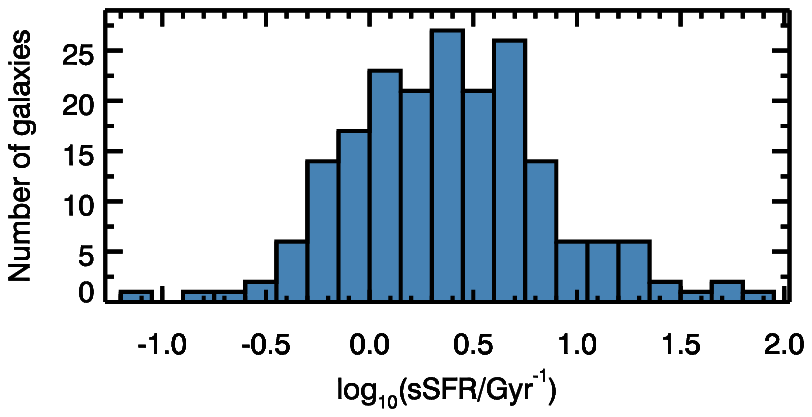}
\caption{The bulk galaxy properties ($M_{\ast}$, SFR, and sSFR) for the KBSS sample used for the analysis presented in this paper. The sample spans over two decades in all three parameters and is representative of the larger KBSS sample, with median M$_{\ast}=9.5\times10^{9}$~M$_{\odot}$, median $\textrm{SFR}=23$~M$_{\odot}$~yr$^{-1}$, and median $\textrm{sSFR}=2.4$~Gyr$^{-1}$. For comparison, the distribution of M$_{\ast}$ for all $z\simeq2-2.7$ KBSS galaxies is shown in grey in the top panel, and a two-sample Kolmogorov-Smirnov (KS) test indicates that the distributions are consistent with one another ($p=0.95$).}
\label{sample_hists}
\end{figure}

Figure~\ref{sample_hists} shows the stellar mass (M$_{\ast}$), star-formation rate (SFR), and specific star-formation rate ($\textrm{sSFR}=\textrm{SFR}/\textrm{M}_{\ast}$) distributions for the paper sample (blue histograms), with the full sample of $z\simeq2-2.7$ KBSS galaxies with M$_{\ast}$ estimates shown for comparison in the top panel (grey histogram). Stellar masses are measured as in \citet{steidel2014} and \citetalias{strom2017}, using the methodology described by \citet{reddy2012}. SFRs are determined using extinction-corrected H$\alpha$ measurements as described by \citetalias{strom2017}, where we have adopted the Galactic extinction curve from \citet{cardelli1989}. Values for M$_{\ast}$, SFR, and sSFR are reported assuming a \citet{chabrier2003} stellar initial mass function (IMF). The sample considered here spans a large range in all three bulk galaxy properties, with median M$_{\ast}=9.5\times10^{9}$~M$_{\odot}$, median $\textrm{SFR}=23$~M$_{\odot}$~yr$^{-1}$, and median $\textrm{sSFR}=2.4$~Gyr$^{-1}$. These values are consistent with the SFR-M$_{\ast}$ relation for galaxies at $2.0<z<2.5$ reported by \citet{whitaker2014}. However, both the \citet{whitaker2014} relation and the median SFR for the KBSS sample discussed here are somewhat higher than would be predicted by the relation from \citet{shivaei2015}, based on a different sample of $z\sim2$ galaxies\footnote{The MOSDEF sample used by \citet{shivaei2015} was selected in a different manner from KBSS, which may account for this difference.}.

\section{Photoionization Model Method}
\label{model_section}

\subsection{Model Grid}
\label{grid_section}

As in \citetalias{steidel2016} and \citetalias{strom2017}, we use Cloudy \citep[v13.02,][]{ferland2013} to predict the nebular spectrum originating from gas irradiated by a given stellar population's radiation field, with the ultimate goal of generating predicted emission line fluxes as a function of \textit{all} of the physical parameters of interest, $f_\textrm{line}(Z_{\ast}, U, Z_{\textrm{neb}}, \textrm{N/O})$. In all cases, we adopt a plane parallel geometry and $n_H=300$~cm$^{-3}$, with the latter motivated by measurements of electron density ($n_e$) in $\langle z\rangle = 2.3$ KBSS galaxies (\citetalias{steidel2016}, \citetalias{strom2017})\footnote{In ionized gas, $n_H\approx n_e$.}. Dust grains are included assuming the ``Orion'' mixture provided as part of Cloudy, with a dust-to-gas ratio that scales linearly with the metallicity of the gas, $Z_{\textrm{neb}}$.

We use stellar population synthesis models from ``Binary Population and Spectral Synthesis'' \citep[BPASSv2;][]{stanway2016,eldridge2017} as the input ionizing spectra, which set the \textit{shape} of the ionizing radiation field. To capture the range of stellar populations that may exist in the parent sample of high-$z$ star-forming galaxies, we employ models from BPASSv2 with constant star-formation histories and varying stellar metallicities, $Z_{\ast}=[0.001,0.002,0.003,0.004,0.006,0.008,0.010,0.014]$ ($Z_{\ast}/Z_{\odot}\approx0.07-1.00$). We use a single age of 100 Myr for all stellar population models, consistent with the characteristic ages of star-forming galaxies at $z\sim2$ \citep[e.g.][]{reddy2012}. The assumption of a fixed age does not significantly impact the photoionization model predictions, as the shape of the UV ionizing SED produced by BPASSv2 models with constant star formation histories reaches equilibrium after $20-30$~Myr. Variations in the \textit{intensity} of the radiation field are parameterized by the ionization parameter $U$($= n_\gamma/n_H$, the dimensionless ratio of the number density of incident ionizing photons to the number density of neutral hydrogen). Because we adopt a single $n_H$, differences in $U$ largely reflect differences in the normalization of the radiation field.

The metallicity of the gas, $Z_{\textrm{neb}}$, is allowed to vary independently of $Z_{\ast}$ and spans $Z_{\textrm{neb}}/Z_{\odot}=0.1-2.0$. As discussed in greater detail elsewhere \citepalias[including][]{steidel2016,strom2017}, $Z_{\textrm{neb}}$ primarily reflects the abundance of O (which is the most abundant heavy element in \ion{H}{2} regions, thus regulating gas cooling and many features of the nebular spectrum). In contrast, $Z_{\ast}$ traces Fe abundance, which provides the majority of the opacity in stellar atmospheres and is critical for determining the details of the stellar wind and mass loss. While we expect the gas and stars to have the same O/H and Fe/H as one another, decoupling $Z_{\textrm{neb}}$ and $Z_{\ast}$ allows us to explicitly test for the presence of non-solar O/Fe in high-$z$ galaxies. Given the high sSFRs observed in these galaxies and young inferred ages, enrichment from core-collapse supernovae (CCSNe) will dominate relative to contributions from Type Ia SNe, making it likely that super-solar O/Fe is in fact more typical than (O/Fe)$_{\odot}$ at high redshift. In the future, we look forward to incorporating stellar population synthesis models with non-solar values of O/Fe---thus allowing a single value of $Z$ to be adopted for both the gas and stars---but without access to such models at the current time, varying $Z_{\textrm{neb}}$ and $Z_{\ast}$ independently is the simplest way to mimic the super-solar O/Fe required to match the observations of typical high-$z$ galaxies' nebular spectra \citepalias{steidel2016}.

Finally, we allow N/O to vary independently of both $Z_{\textrm{neb}}$ and $Z_{\ast}$, with a minimum $\log(\textrm{N/O})=-1.8$ ($[\textrm{N/O}] = -0.94$). In the local universe, the N/O measured in \ion{H}{2} regions and star-forming galaxies is known to correlate with the overall O/H \citep[e.g.,][]{vanzee1998}, exhibiting a low, constant value ($\log(\textrm{N/O})\approx-1.5$) at $12+\log(\textrm{O/H})\lesssim8.0$ and increasing with increasing O/H past this critical level of enrichment. Although the nucleosynthetic origin of N remains largely uncertain, this behavior is often interpreted as a transition between ``primary'' and ``secondary'' production of N \citep[e.g.,][]{edmunds1978,vilacostas1993}. Notably, among a compilation of local extragalactic \ion{H}{2} regions from \citet[][hereafter Pil12]{pilyugin2012}, there is a factor of $\sim2-3$ scatter in N/O at \textit{fixed} O/H, particularly near the transition metallicity ($12+\log(\textrm{O/H})\approx8.0$). Methods 	that adopt (explicitly or implicitly) a single relation with no scatter between N/O and O/H marginalize over this intrinsic physical scatter, which affects the accuracy of the inferred O/H. Some methods, like HII-CHI-MISTRY \citep{perez-montero2014}, do allow for some scatter in N/O at fixed O/H, but still impose a prior that prevents any enrichment in N/O greater than that observed in the calibration sample. Because we are interested in quantifying the relationship between N/O and O/H among high-$z$ galaxies, including the degree of intrinsic scatter, we opt to allow any value of $\log(\textrm{N/O})\geq-1.8$.

Although varying other physical conditions (including $n_H$) can also affect the resulting nebular spectrum, the effect of such differences within the range observed in the KBSS sample (where most galaxies are consistent with $n_e=300$~cm$^{-3}$) is considerably smaller than the effects of altering the shape and/or normalization of the ionizing radiation field (i.e., varying $Z_{\ast}$ and/or $U$) or the chemical abundance pattern in the gas (i.e., varying $Z_{\textrm{neb}}$ or N/O). Thus, in summary, we assemble a four-dimensional photoionization model grid, spanning the following parameter space:
\begin{eqnarray}
&&Z_{\ast}/Z_{\odot} \approx [0.07,1.00] \textrm{, from BPASSv2} \nonumber \\
&&Z_{\textrm{neb}}/Z_{\odot} = [0.1,2.0] \textrm{, every 0.1 dex} \nonumber \\
&&\log(U) = \nonumber [-3.5,-1.5] \textrm{, every 0.1 dex} \\
&&\log(\textrm{N/O}) \geq -1.8. \nonumber
\end{eqnarray}

\begin{figure*}
\centering
\includegraphics[height=0.42\textheight]{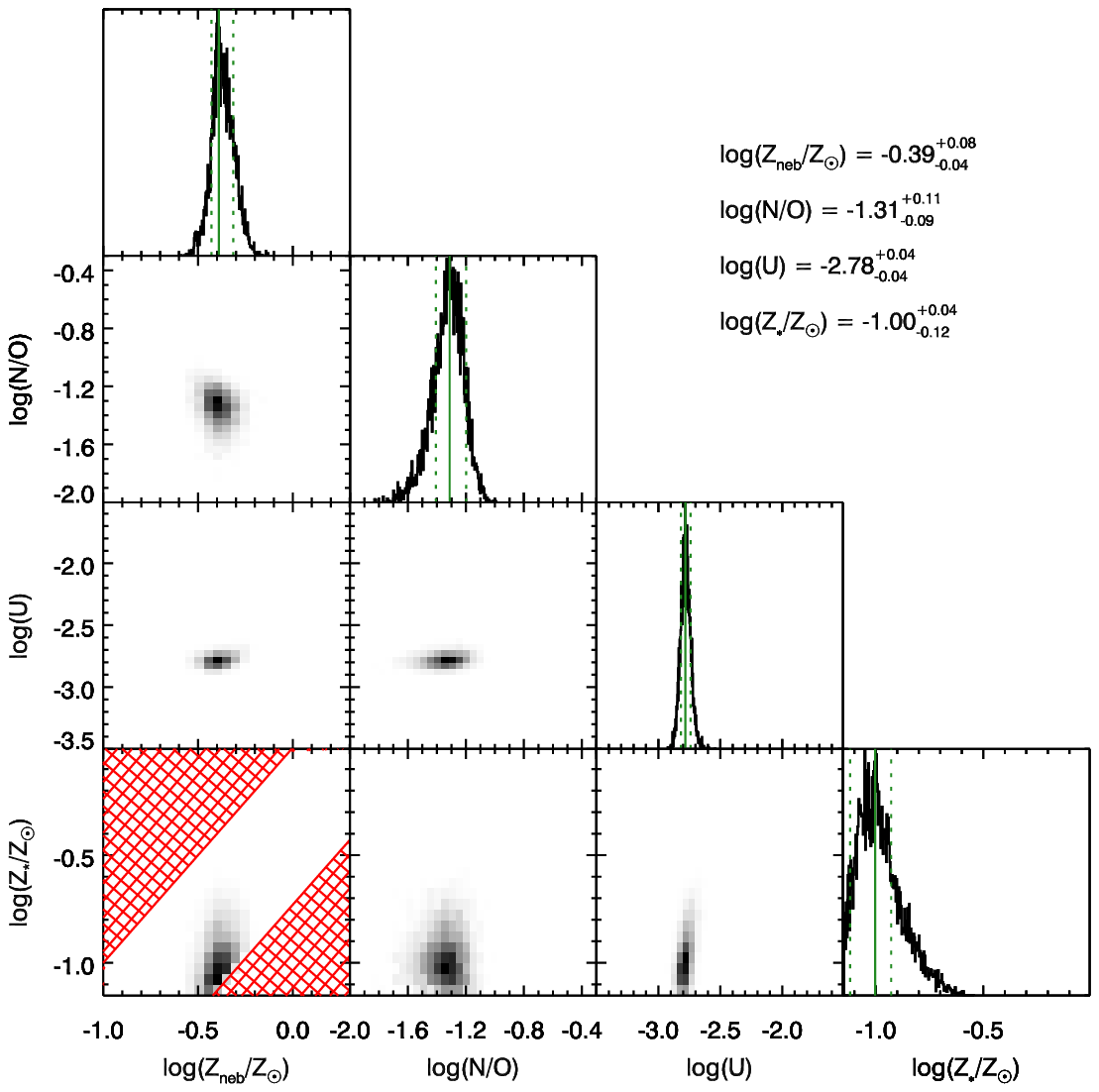}
\vspace{-0.1in}
\caption[MCMC results for Q1442-BX160]{The posterior PDFs for log($Z_{\textrm{neb}}/Z_{\odot}$), log(N/O), log($U$), and log($Z_{\ast}/Z_{\odot}$) for Q1442-BX160, as estimated from the MCMC routine. The displayed range reflects the bounds on the parameters, which are listed in the text. The red shaded regions in the panel showing the 2D pairwise posterior for log($Z_{\ast}/Z_{\odot}$) and log($Z_{\textrm{neb}}/Z_{\odot}$) reflect disallowed regions of parameter space with $\log(Z_{\textrm{neb}}/Z_{\ast})<0.0$ or $\log(Z_{\textrm{neb}}/Z_{\ast})>0.73$. The solid green lines show the locations of the MAP estimates for the marginalized posteriors, with the dashed green lines reflecting the 68\% highest posterior density interval. The MAP estimates and the corresponding errors for each parameter are printed above the figure. Q1442-BX160 is an example of an object with bound, single-peaked posteriors in all four parameters, where the parameter estimation is relatively straightforward.}
\label{good_mcmc}
\centering
\includegraphics[height=0.42\textheight]{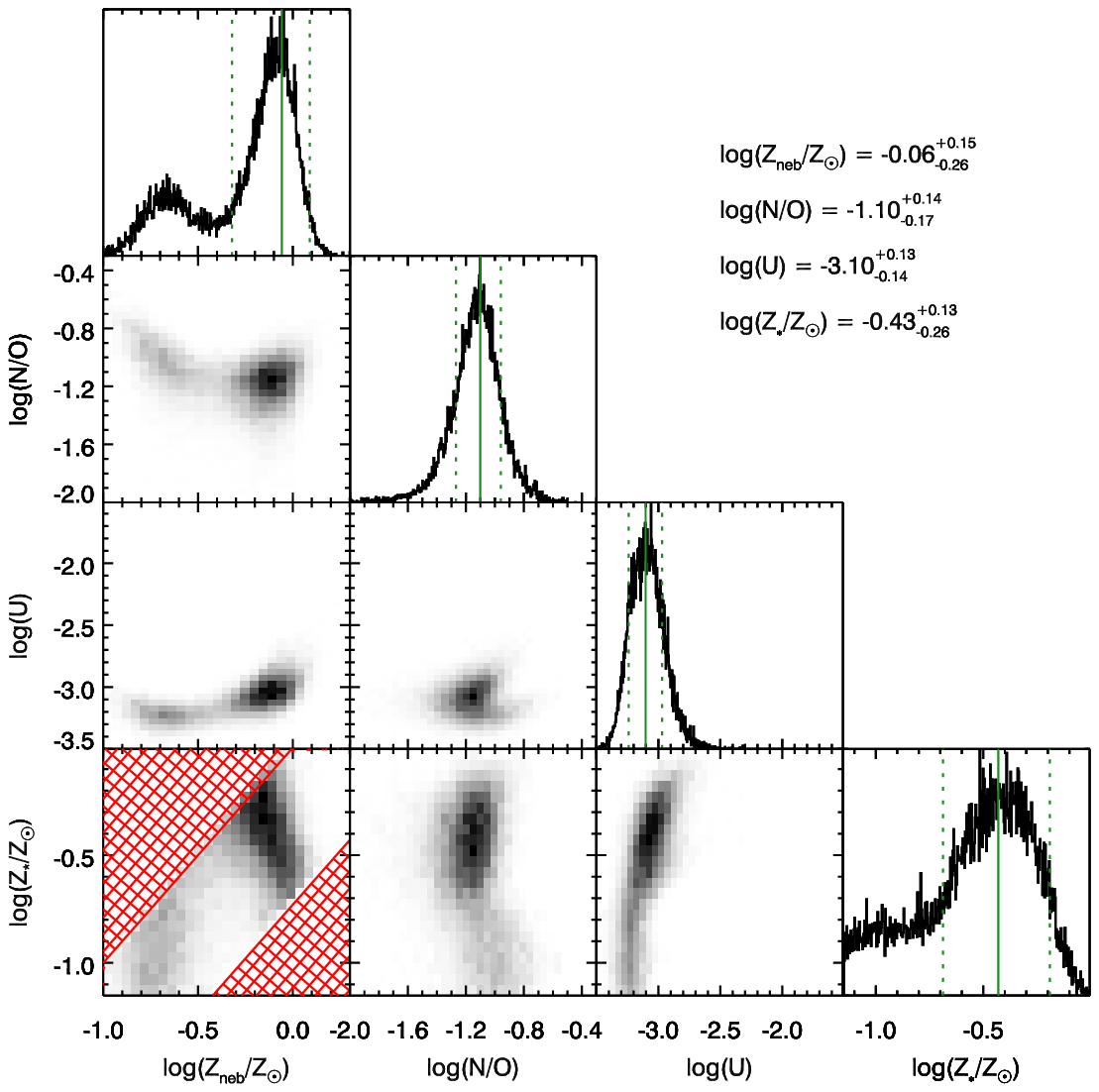}
\vspace{-0.1in}
\caption[MCMC results for Q1009-BX218]{The posterior PDFs for Q1009-BX218, shown in the same manner as Figure~\ref{good_mcmc}. Q1009-BX218 highlights the case where the multi-dimensional posterior is double-peaked, because of degeneracies in the photoionization model grid. This is reflected in both the 2D posteriors and marginalized posteriors for single parameters. Nevertheless, because one solution is significantly more likely, it is possible to infer log($Z_{\textrm{neb}}/Z_{\odot}$), log(N/O), log($U$), and log($Z_{\ast}/Z_{\odot}$), as described in the text and shown by the solid green lines.}
\label{bad_mcmc}
\end{figure*}

\subsection{MCMC Method}

We do not know \textit{a priori} the degeneracies among $Z_{\ast}$, $U$, $Z_{\textrm{neb}}$, and N/O, in terms of determining the observed nebular spectra of high-$z$ galaxies, and it is possible that the correlations found in low-$z$ samples may not hold at higher redshifts. Indeed, two of the main goals of this work are to: (1) quantitatively determine these parameters in individual high-$z$ galaxies, discussed in this section, and (2) evaluate the presence of any intrinsic correlations between them among the high-$z$ galaxy population, discussed in Section~\ref{correlation_section}. Given the relatively large size of the total parameter space outlined above and the expectation that the posterior probability density functions (PDFs) for the parameters may not be normally distributed, we employ a Markov Chain Monte Carlo (MCMC) technique. Such a method allows us to efficiently determine the combinations of $Z_{\ast}$, $U$, $Z_{\textrm{neb}}$, and N/O that are consistent with the entire observed rest-optical nebular spectra of individual objects, as well as to quantify the correlations between inferred parameters.

We initialize the chain at the photoionization model grid point that best matches the combination of emission lines measured from a galaxy's spectrum. The MCMC sampler then explores log parameter space, using a normal proposal algorithm whose width is adjusted based on the acceptance rate to allow for more efficient local sampling. In general, flat priors are adopted within the boundaries of the parent photoionization model grid, with one additional constraint:
$$
0.0 < \log(Z_{\textrm{neb}}/Z_{\ast}) < 0.73.
$$
Given our assumption that $Z_{\textrm{neb}}$ traces O/H and $Z_{\ast}$ traces Fe/H, limits on $\log(Z_{\textrm{neb}}/Z_{\ast})$ in our model reflect limits on [O/Fe], which are informed by stellar nucleosynthesis and galactic chemical evolution models. The upper limit we impose on $\log(Z_{\textrm{neb}}/Z_{\ast})$ corresponds to the highest [O/Fe] expected for the Salpeter IMF-averaged yields from Fe-poor ($Z_{\ast}=0.001$) CCSNe \citep{nomoto2006}. The lower limit is equivalent to $[\textrm{O/Fe}]=0.0$, which is the default implicitly assumed by most stellar population synthesis models and, by extension, most photoionization model methods for determining physical conditions in galaxies. Since high-$z$ galaxies are typically young and rapidly star-forming, the true value of [O/Fe]---and, thus, $\log(Z_{\textrm{neb}}/Z_{\ast})$---likely falls in this range for most objects.

If the proposed step is not coincident with a grid point in the parent photoionization model grid, the predicted emission line fluxes are calculated via trilinear interpolation of the grid in $Z_{\ast}$, $U$, and $Z_{\textrm{neb}}$ and subsequently scaling the model [\ion{N}{2}]$\lambda6585$ flux based on the proposed value of log(N/O). For example, $\log(\textrm{N/O})=-0.86$ ($[\textrm{N/O}]=0$) corresponds to a scale factor of 1, whereas $\log(\textrm{N/O})=-1.5$ ($[\textrm{N/O}]=-0.64$) corresponds to a scale factor of 0.23 relative to the default [\ion{N}{2}]$\lambda6585$ flux from our parent photoionization model grid, which assumes solar N/O\footnote{Assuming a constant value of N/O in Cloudy does not have a significant impact on the structure of the nebula.}.

Because the photoionization model line intensities are reported relative to H$\beta$, we convert the observed line fluxes and errors onto the same scale by first correcting all of the line measurements for differential reddening due to dust and then dividing by the flux in H$\beta$. The final errors used in the MCMC analysis include the contribution from the error on the Balmer decrement, but do not account for systematic uncertainties in the choice of extinction curve. In all cases, the line-of-sight extinction curve from \citet{cardelli1989} is used to correct the line fluxes, but because many commonly-used extinction curves have a similar shape at rest-optical wavelengths, our results do not change significantly if we adopt an SMC-like extinction curve. Finally, although a Balmer decrement of 2.86 is widely adopted as the fiducial value for Case B recombination \citep{osterbrock1989}, the exact value is sensitive to both $n_e$ and the electron temperature, $T_e$ (and, thus, the shape of the ionizing radiation field). To remain self-consistent, we adopt a Case B value for the Balmer decrement based on the predicted H$\alpha$ flux (in units of H$\beta$) at every proposed step, which ranges from $2.85-3.14$ in the parent photoionization model grid.

Our MCMC method is able to account for low-SNR measurements of emission lines in the same manner as more significant detections, and thus no limits are used. If a measured line flux is formally negative \citepalias[as measured using the custom spectral analysis tool \textit{MOSPEC}, described by][]{strom2017}, the expectation value for the line flux is assumed to be zero, with a corresponding error $\sigma_{f/\textrm{H}\beta} = \frac{\sigma_{f}}{f_{\textrm{H}\beta}}$.

Convergence of the MCMC to the posterior is evaluated using the potential scale reduction factor, $\hat{R}$ \citep[also known as the Gelman-Rubin diagnostic,][]{gelman1992}, using equal-length segments of the chain to compare the dispersion within the chain to the dispersion between portions of the chain. When the chain segments look very similar to one another, $\hat{R}$ will trend from higher values toward 1. When $\hat{R}<1.05$ for all four model parameters, we assume approximate convergence has been achieved\footnote{In two cases (Q1549-BX197 and Q2343-BX505), $\hat{R}$ remains above 1.05 even after a large number of steps. This can occur when there are multiple, well-separated peaks in the posterior distribution. Because both peaks were well-sampled despite large values of $\hat{R}$ for the total chain, we include the galaxies in the final sample.}.

\section{Measurements of the Chemistry and Ionization in Individual $z\sim2$ Galaxies}
\label{results_section}

\begin{deluxetable}{llll}
\tablecaption{MCMC results for $z\simeq2-2.7$ KBSS-MOSFIRE galaxies}
\tablehead{
Sample & $N_{\textrm{gal}}$
}
\startdata
Galaxies used in the photoionization model analysis & 202 \\
Galaxies with bound PDFs for all parameters & 121 \\
Galaxies with bound PDFs for all but N/O & 27 \\
\\
Galaxies with bound $\log(Z_{\textrm{neb}}/Z_{\odot})$ PDF & 163 \\
Galaxies with bound $\log$(N/O) PDF & 164 \\
Galaxies with bound $\log(U)$ PDF & 193 \\
Galaxies with bound $\log(Z_{\ast}/Z_{\odot})$ PDF & 169 \\
\\
Galaxies with 2+ peaks in $\log(Z_{\textrm{neb}}/Z_{\odot})$ & 88 \\
Galaxies with 2+ peaks in $\log$(N/O) & 35 \\
Galaxies with 2+ peaks in $\log(U)$ & 24 \\
Galaxies with 2+ peaks in $\log(Z_{\ast}/Z_{\odot})$ & 111 \\
\\
Galaxies with an upper limit on $\log$(N/O) &  37 \\
Galaxies with an upper limit on $\log(U)$ & 1 \\
Galaxies with an upper limit on $\log(Z_{\ast}/Z_{\odot})$ & 15 \\
Galaxies with a lower limit on $\log(Z_{\ast}/Z_{\odot})$ & 2
\enddata
\tablecomments{A marginalized PDF is considered bound when the 68\% highest posterior density interval (1) does not bracket a local minimum with lower probability than the probability at the interval boundaries and (2) does not coincide the edge of the sampled range in the parameter.}
\label{sample_table}
\end{deluxetable}

Figures~\ref{good_mcmc} and \ref{bad_mcmc} show the MCMC results for two individual galaxies in the KBSS sample, highlighting the range of possible outcomes. The results for Q1442-BX160 contain a single preferred solution, with pairwise 2D posteriors that resemble bivariate Gaussian distributions. In contrast, the joint posterior PDF for Q1009-BX218 reveals two distinct solutions and notably non-normal marginalized posteriors.

Nearly two-thirds of objects ($\sim65$\%), including Q1009-BX218, have 2 or more peaks in the marginalized posterior for at least one model parameter. Nevertheless, the distribution of probability between multiple peaks still allows for a single solution to be preferred in many cases. This is true for Q1009-BX218, where the 68\% highest density interval (HDI, the narrowest range that includes 68\% of the distribution) contains a single dominant peak in log($Z_{\textrm{neb}}/Z_{\odot}$).

We adopt the following criteria for identifying ``bound'' marginalized posteriors, which are those posterior distributions from which a preferred model parameter can be estimated: the 68\% HDI must contain either a single peak or, if 2 or more peaks are present, there cannot exist a local minimum in the posterior PDF with lower probability than the probability at the edge of the 68\% HDI. In addition, the 68\% HDI must not abut the edge of the model grid. For bound posteriors, we adopt the maximum \textit{a posteriori} (MAP) estimate as the inferred value for the model parameter. The MAP estimate is taken to be the peak value of the Gaussian-smoothed histogram representing the marginalized posterior. The asymmetric errors on this estimate are determined by calculating the 68\% HDI for the same posterior. In total, 121 galaxies have bound posteriors and, thus, MAP estimates for log($Z_{\textrm{neb}}/Z_{\odot}$), log(N/O), log($U$), and log($Z_{\ast}/Z_{\odot}$).

For cases where one edge of the 68\% HDI does in fact coincide with the edge of the parent photoionization model grid for a given parameter, we assume that the posterior reflects a limit and record the opposite boundary of the 68\% HDI as an upper or lower limit in that parameter, corresponding to a $2\sigma$ limit. Table~\ref{sample_table} provides a more detailed summary of the MCMC results, including the number of galaxies with limits on a given parameter.

\begin{figure*}
\centering
\includegraphics{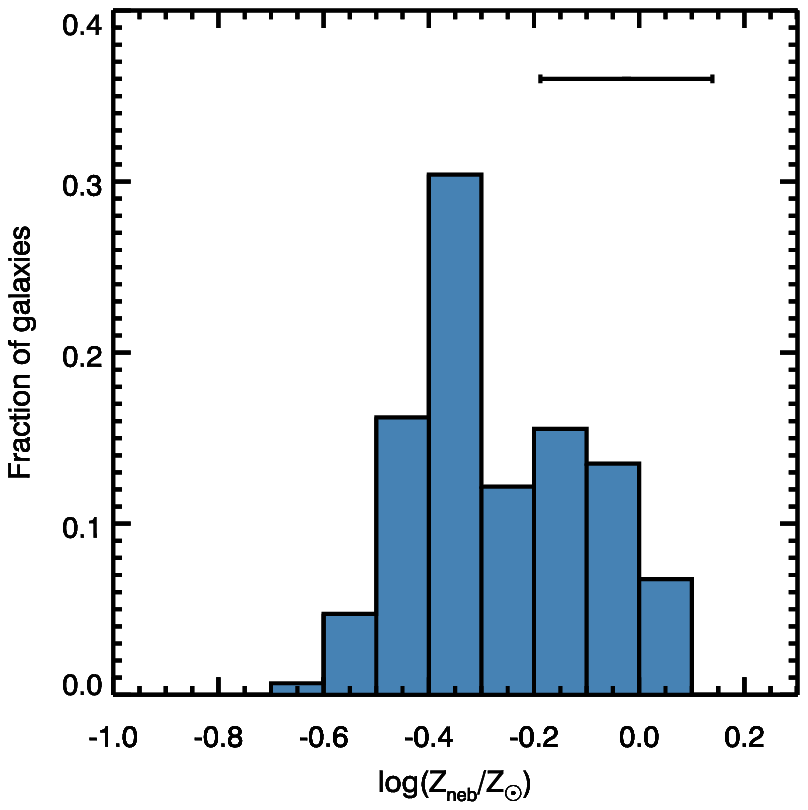}
\includegraphics{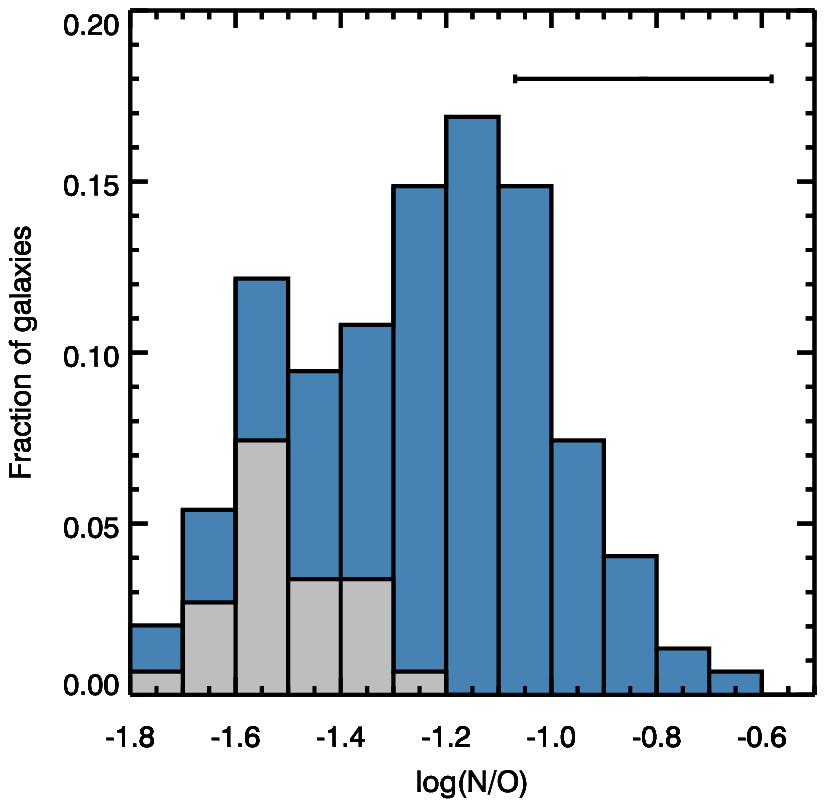}\\
\includegraphics{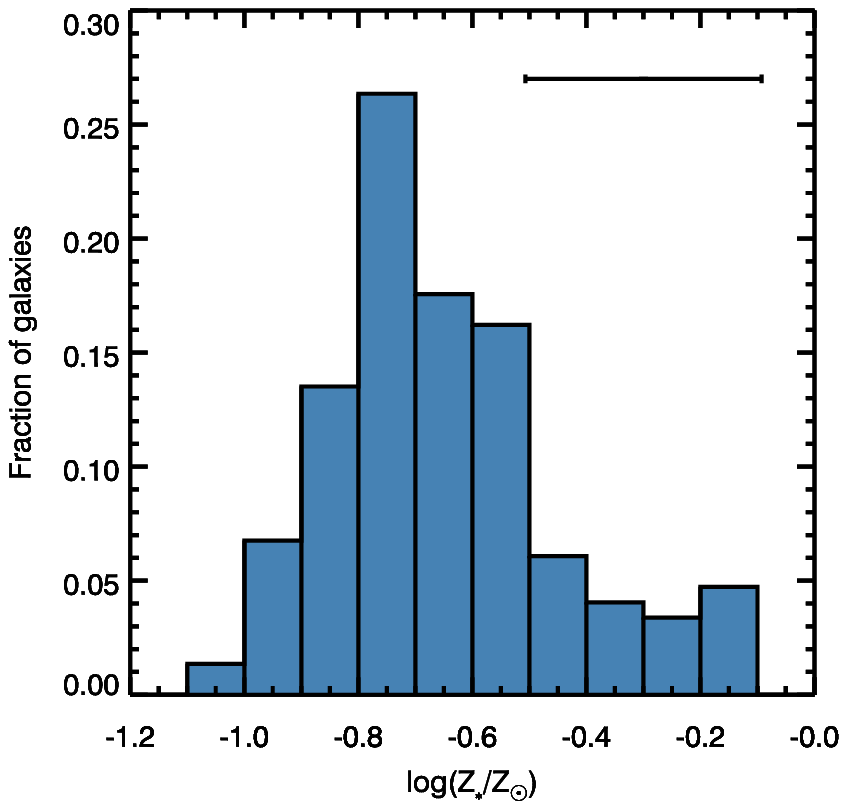}
\includegraphics{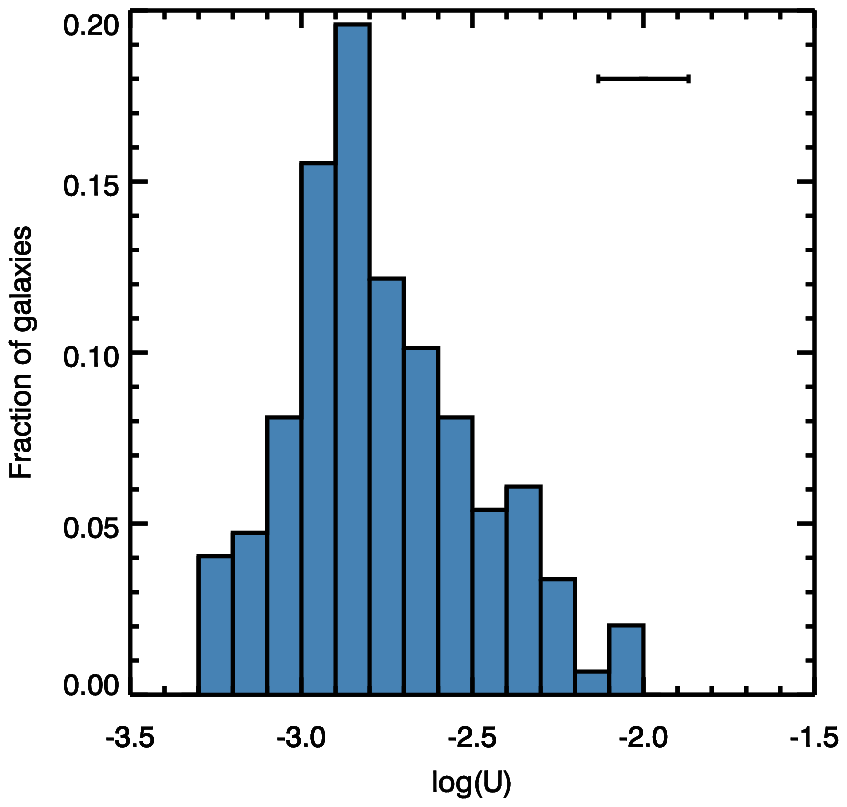}
\caption[Distributions of model-inferred log($Z_{\textrm{neb}}/Z_{\odot}$), log(N/O), log($U$), and log($Z_{\ast}/Z_{\odot}$) for $z\simeq2-2.7$ KBSS galaxies]{Distributions of model-inferred log($Z_{\textrm{neb}}/Z_{\odot}$), log(N/O), log($Z_{\ast}/Z_{\odot}$), and log($U$) for 148 KBSS galaxies, showing for the first time self-consistent estimates of the physical conditions in a large sample of individual high-$z$ galaxies. The contribution from galaxies with upper limits on log(N/O) is shown as the grey portion of the histogram in the upper right panel. For reference, the average error ($\Delta_{68}$) for each parameter is shown in the upper right corner of each panel. These results are consistent in many respects with previous estimates of these parameters, including the range of log($U$) we reported for KBSS galaxies in \citetalias{strom2017}, but offer new insights into the range and distributions of physical characteristics common in $\langle z\rangle = 2.3$ galaxies.}
\label{cloudy_hists}
\end{figure*}

Figure~\ref{cloudy_hists} shows the distributions of inferred log($Z_{\textrm{neb}}/Z_{\odot}$), log(N/O), log($U$), and log($Z_{\ast}/Z_{\odot}$) for the 121 KBSS galaxies with bound posteriors in all four model parameters and the 27 galaxies with upper limits on log(N/O), resulting in a total sample of 148 galaxies. The upper right panel, showing the distribution of log(N/O), separates the contribution from galaxies with N/O limits, illustrated as the grey portion of the histogram; we discuss this subsample in more detail below. The interquartile ranges in the model parameters for this subsample (including those with upper limits on N/O) are
\begin{eqnarray}
\log(Z_{\textrm{neb}}/Z_{\odot})_{50} &=& [-0.40,-0.13] \nonumber \\
\log(\textrm{N/O})_{50} &=& [-1.44,-1.07] \nonumber \\
\log(U)_{50} &=& [-2.93,-2.58] \nonumber \\
\log(Z_{\ast}/Z_{\odot})_{50} &=& [-0.79,-0.53]. \nonumber
\end{eqnarray}
These characteristic ranges are consistent with inferences of these parameters by other means. In particular, the range of interred log($U$) is in good agreement with the range we previously reported for individual KBSS galaxies in \citetalias{strom2017}, using a more limited photoionization model approach. 

\subsection{Multi-peaked Posteriors and Limits}
\label{peaks_and_limits}

\begin{figure*}
\centering
\includegraphics{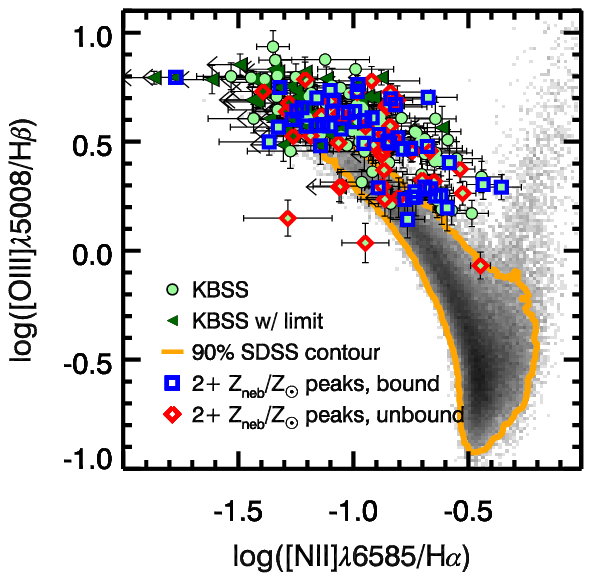}
\hspace{-0.80in}
\includegraphics{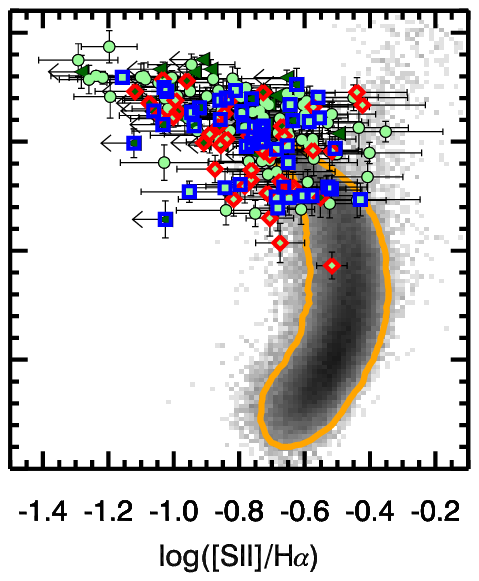}
\includegraphics{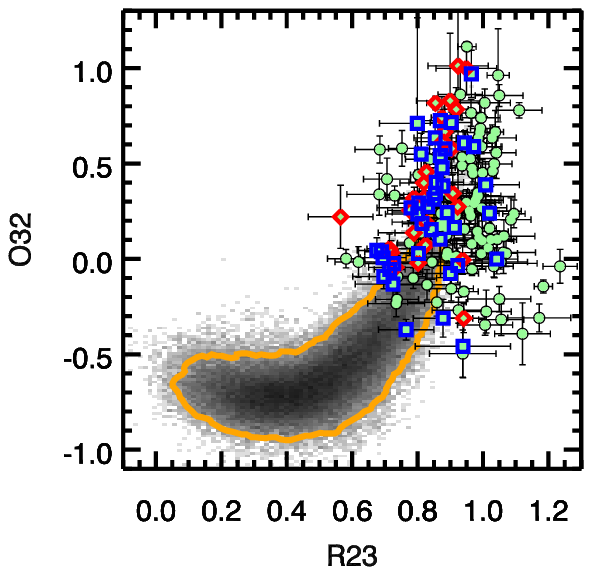}\\
\vspace{-0.7in}
\includegraphics{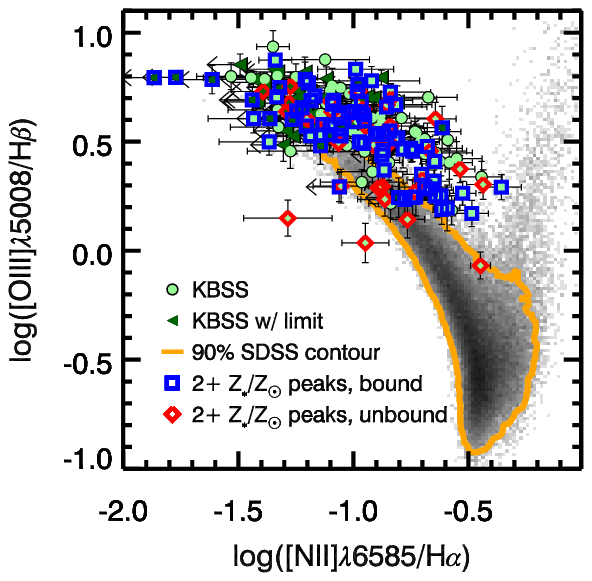}
\hspace{-0.80in}
\includegraphics{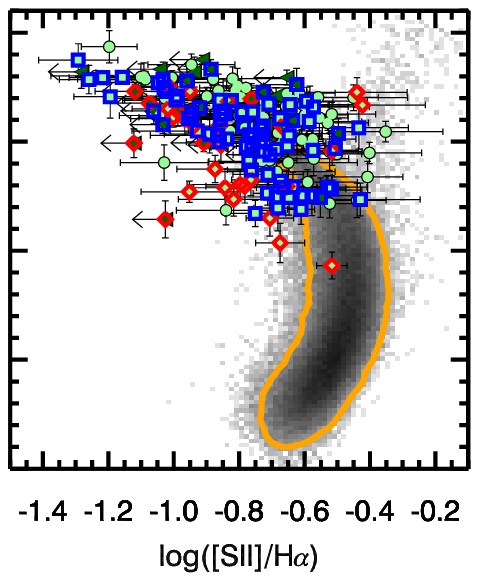}
\includegraphics{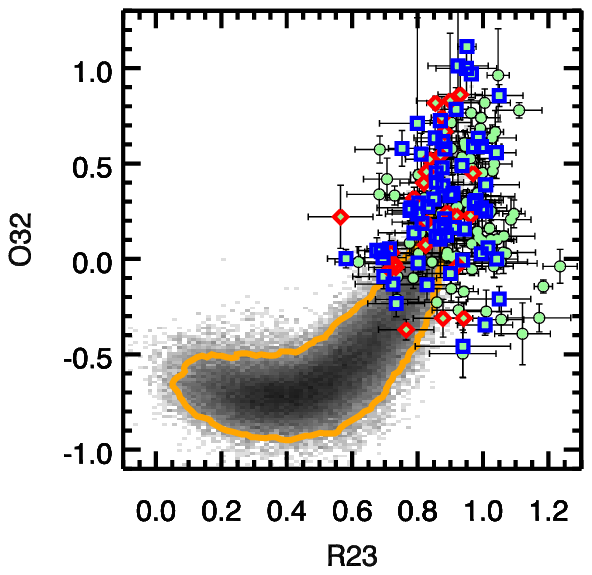}
\caption{Three commonly-used 2D emission line ratio diagrams: the N2-BPT diagram (left column), the S2-BPT diagram (center column), and the dust-corrected O32-R23 diagram (right column). The locus of $z\sim0$ galaxies from SDSS is shown in greyscale in each panel, with an orange contour enclosing 90\% of the sample. The sample of $\langle z\rangle = 2.3$ galaxies from KBSS used in this paper is shown by the green points with error bars; galaxies with $2\sigma$ upper limits on [\ion{N}{2}] and [\ion{S}{2}] are instead illustrated as darker green triangles. The top row shows the location of galaxies with multiple peaks in $\log(Z_{\textrm{neb}}/Z_{\odot})$, separated by whether the posterior is bound (blue squares) or unbound (red diamonds). The bottom row shows the location of galaxies with multiple peaks in $\log(Z_{\ast}/Z_{\odot})$ in the same manner.}
\label{ratio_diagrams}
\end{figure*}

Multipeaked posteriors are most frequently observed for log($Z_{\textrm{neb}}/Z_{\odot}$) and log($Z_{\ast}/Z_{\odot}$). Figure~\ref{ratio_diagrams} shows the location of galaxies with 2 or more peaks in the marginalized posteriors for these parameters in the N2-BPT, S2-BPT, and O32-R23 diagrams, highlighting differences in the spectra of galaxies with unbound posteriors (e.g., those with equal power shared among 2 or more peaks, identified by red diamonds) and those with bound posteriors (e.g., those with a single dominant peak or group of peaks, identified by blue squares). A few trends are common for both parameters, including the clustering on the left (low-R23) side of the galaxy locus in the O32-R23 diagram (shown in the right column). This region of parameter space is populated by photoionization model grid points with either low \textit{or} high $Z_{\textrm{neb}}$, whereas higher values of R23 at fixed O32 can only be achieved by combining moderate $Z_{\textrm{neb}}$ with low $Z_{\ast}$.

Interestingly, galaxies with multi-peaked posteriors in log($Z_{\textrm{neb}}/Z_{\odot}$) or log($Z_{\ast}/Z_{\odot}$) are well-mixed with the total sample in the N2-BPT diagram (left column in Figure~\ref{ratio_diagrams}), regardless of whether the posteriors are bound. The same is true for galaxies with bound multi-peaked posteriors in either parameter in the S2-BPT diagram (blue squares in the center column). However, galaxies with unbound multi-peaked posteriors in log($Z_{\ast}/Z_{\odot}$) almost exclusively occupy a region of parameter space to the lower left of the ridge-line of KBSS galaxies in the S2-BPT diagram (the red diamonds with low [\ion{O}{3}]/H$\beta$ at fixed [\ion{S}{2}]/H$\alpha$ in the bottom center panel of Figure~\ref{ratio_diagrams}). In general, these galaxies do not exhibit multiple isolated peaks in log($Z_{\ast}/Z_{\odot}$), but rather relatively flat posteriors with 2 or more local maxima spread over a range in log($Z_{\ast}/Z_{\odot}$). Galaxies in this region of parameter space, as with those found at low-R23 and high-O32, are consistent with a range in $Z_{\ast}$, and additional information is required to break the degeneracy between possible solutions. In the future, we hope to incorporate constraints on the detailed stellar photospheric absorption features observed in the non-ionizing UV spectra of individual galaxies in KBSS, but note that such data are not always available when trying to infer physical conditions from galaxies' rest-optical spectra (either from KBSS or other surveys).

There are 27 galaxies with marginalized posterior PDFs that reflect upper limits on log(N/O) while having bound posteriors for the other three parameters. These galaxies are not preferentially distributed in a specific location in the N2-BPT, S2-BPT, and O32-R23 diagrams, but all such galaxies have $\textrm{SNR}<2$ measurements of [\ion{N}{2}]$\lambda6585$, which was the significance threshold we adopted in \citetalias{strom2017}. In contrast, only $\sim18$\% of galaxies with bound posteriors in all four model parameters, including log(N/O), have $\textrm{SNR}<2$ in their [\ion{N}{2}]$\lambda6585$ measurements. Unsurprisingly, a two-sample Kolmogorov-Smirnov (KS) test indicates that the distributions of [\ion{N}{2}]$\lambda6585$ SNR for the two subsamples are significantly unlikely to be drawn from the same parent population. This result follows naturally from the fact that [\ion{N}{2}] is the only available constraint on N/O. Because the likelihood of having an upper limit on log(N/O) results almost exclusively from a low SNR [\ion{N}{2}]$\lambda6585$ measurement, which (as we show later in Section~\ref{snr_section} and Table~\ref{corr_w_observables}) has only a minor impact on the other model parameters, we choose to include galaxies with N/O limits but bound posteriors for the other three model parameters in our analysis.

\subsection{Inferences Using Only ``BPT" Lines}

Of the 202 total galaxies in our sample, 27 do not have measurements of or limits on the [\ion{O}{2}]$\lambda\lambda3727,3729$ doublet or the [\ion{Ne}{3}]$\lambda3869$ line, which are observed in $J$-band for galaxies at these redshifts. To understand the effect that using only lines present in $H$- and $K$-band (i.e., the ``BPT'' lines: H$\alpha$, H$\beta$, [\ion{O}{3}]$\lambda5008$, [\ion{N}{2}]$\lambda6585$, and [\ion{S}{2}]$\lambda\lambda6718,6732$) might have on the MCMC results, we compare the model results for the 109 galaxies with spectral coverage of all of the available emission lines and bound PDFs for all parameters with the results for the same galaxies when [\ion{O}{2}]$\lambda\lambda3727,3729$ and [\ion{Ne}{3}]$\lambda3869$ are excluded. Of these, 66 galaxies also have bound PDFs for all parameters when only considering the BPT lines. In general, differences in the model parameters are small---$\sim0.01-0.03$~dex for log($Z_{\textrm{neb}}/Z_{\odot}$), log(N/O), and log($U$) and $\sim0.06$~dex for log($Z_{\ast}/Z_{\odot}$)---especially when compared to the median uncertainty on the parameter estimates ($\sim0.1-0.2$~dex, when all the available emission lines are considered). The uncertainties are larger by $\sim0.05$~dex when only the BPT lines are used in the MCMC method. This comparison demonstrates the importance of using more emission line measurements to achieve more precise results, but confirms that no large systematic errors are introduced when only the BPT lines are available.

\subsection{The LM1 Composite Spectrum}

\begin{figure*}
\centering
\includegraphics{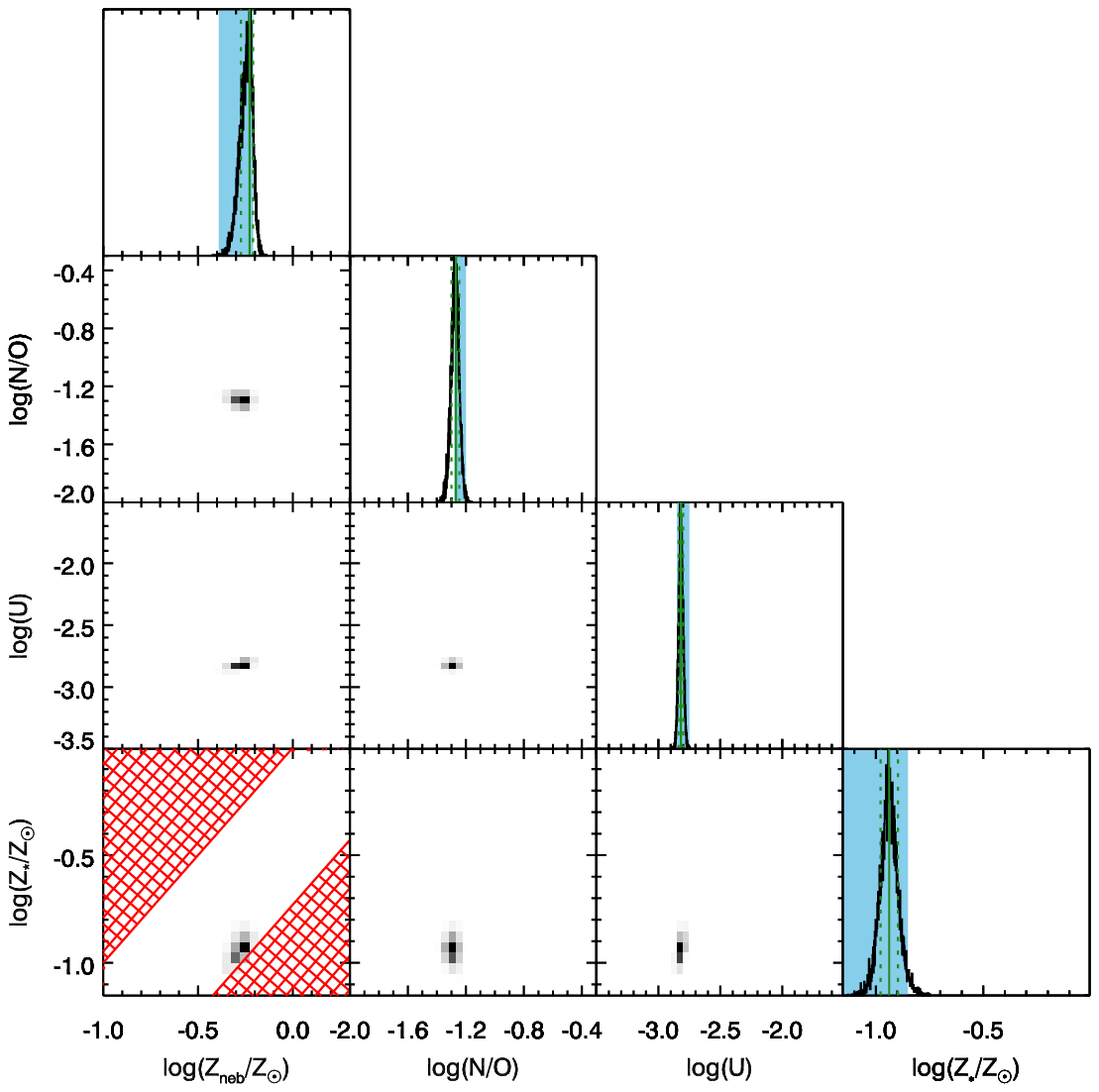}
\caption[MCMC results for the composite spectrum from \citetalias{steidel2016}]{The MCMC results for the LM1 composite spectrum, shown in the same manner as Figure~\ref{good_mcmc}. The blue shaded regions indicate the values for each quantity and the corresponding uncertainties reported in \citetalias{steidel2016}; see also Table~\ref{lm1_table}.}
\label{lm1_mcmc}
\end{figure*}

We also test our method using measurements from the rest-optical composite spectrum presented in \citetalias{steidel2016}, where we were able to estimate O/H, N/O, $U$, and $Z_{\ast}$ using a number of independent cross-checks. The ``LM1'' composite from that analysis was constructed using MOSFIRE observations of 30 KBSS galaxies with $z=2.40\pm0.12$. As described in \citetalias{steidel2016}, the $J$, $H$, and $K$-band spectra of the galaxies were corrected for differential slit losses between bands and shifted into the rest-frame according to the measured nebular redshift, with the normalization of each galaxy's spectra adjusted to account for small redshift differences among the sample. The final composite spectrum is a straight average of the individual shifted and scaled spectra, excluding regions near bright OH sky lines.

To make a fair comparison with the individual galaxies in our sample, for which only the strongest rest-optical nebular emission lines are typically measurable, we use the measurements of [\ion{O}{2}]$\lambda\lambda3727,3729$, [\ion{Ne}{3}]$\lambda3869$, H$\beta$, [\ion{O}{3}]$\lambda5008$, H$\alpha$, [\ion{N}{2}]$\lambda6585$, and [\ion{S}{2}]$\lambda\lambda6718,6732$ reported in \citetalias{steidel2016}. Although we reported a limit on [\ion{O}{3}]$\lambda4364$ for the LM1 composite in \citetalias{steidel2016}, we choose not to include it here, as it was not observed in the spectrum of each individual galaxy in the LM1 sample and is generally not available for individual high-$z$ galaxies. The MCMC results using only the strong-line measurements are shown in Figure~\ref{lm1_mcmc} and show narrow and well-defined marginalized posteriors for each of the four model parameters.

The parameter estimates for the LM1 composite from our photoionization model method are very similar to the values from \citetalias{steidel2016} (identified by shaded blue regions in Figure~\ref{lm1_mcmc}, see also Table~\ref{lm1_table}). In that paper, we were able to directly incorporate information about the photospheric absorption features observed in the non-ionizing UV composite spectrum in order to infer $Z_{\ast}$, and we also leveraged measurements of the auroral [\ion{O}{3}]$\lambda\lambda1661,1666$ lines in the rest-UV as an independent constraint on $Z_{\textrm{neb}}$. In addition, the N/O inferred for the LM1 composite using two locally-calibrated empirical strong-line calibrations was found to be in excellent agreement with the N/O determined through comparisons of the LM1 measurements with photoionization models alone. Given the internal consistency of the analysis presented in \citetalias{steidel2016}, the agreement between the new photoionization model method results and the earlier estimates is reassuring.

The inferred values of log($Z_{\textrm{neb}}/Z_{\odot}$), log(N/O), and log($U$) for the LM1 composite spectrum discussed above (Table~\ref{lm1_table}) are close to the most common values reported for individual galaxies in Figure~\ref{cloudy_hists}.  At the same time, the model stellar metallicity inferred for the LM1 composite ($Z_{\ast}/Z_{\odot}=0.12$) is somewhat lower than the median value for individual galaxies ($Z_{\ast}/Z_{\odot}=0.20$). Although it is possible that the galaxies used to construct the LM1 composite are not characteristic of the full range in $Z_{\ast}$ present in the entire $z\simeq2-2.7$ KBSS sample, it is important to remember that the constraints on $Z_{\ast}$ from the MCMC method presented in this paper are indirect and based on the ionizing spectral shape of the input stellar population, as reflected by the ratios of the rest-optical emission lines. In \citetalias{steidel2016}, we were able to leverage direct observations of photospheric line-blanketing at non-ionizing UV wavelengths to measure $Z_{\ast}$. That the two analyses agree for the LM1 composite suggests that the BPASS models near $Z_{\ast}/Z_{\odot}\sim0.1$ are self-consistent in terms of their ionizing radiation fields and photospheric absorption---but the overall mapping of radiation field hardness to $Z_{\ast}$ remains unavoidably model-dependent. Extending the kind of analysis presented in \citetalias{steidel2016} and in this paper either to individual galaxies or to composite spectra in bins of photoionization model-inferred $Z_{\ast}$ could therefore provide a method for cross-checking the consistency of stellar population models across a range of metallicities.

\begin{deluxetable}{lrr}
\tablecaption{Results for the LM1 Composite}
\tablehead{
Quantity & \multicolumn{1}{c}{\citet{steidel2016}} & \multicolumn{1}{c}{this paper}}
\startdata
$Z_{\textrm{neb}}/Z_{\odot}$ & $0.50\pm0.10$ & $0.59^{+0.03}_{-0.06}$\\
log(N/O) & $-1.24\pm0.04$ & $-1.27^{+0.02}_{-0.03}$\\
log($U$) & $-2.8$ & $-2.82\pm0.02$\\
$Z_{\ast}/Z_{\odot}$ & $\simeq0.1$ & $0.12\pm0.01$
\enddata
\label{lm1_table}
\end{deluxetable}

\subsection{Trends with Galaxy Spectral Properties}

\subsubsection{Location in Line-ratio Diagrams}

\begin{figure*}
\centering
\hspace{-0.1in}
\includegraphics{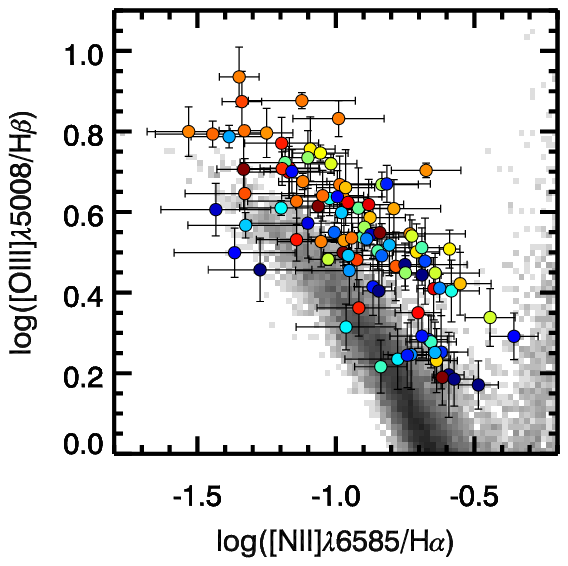}
\hspace{-1.35in}
\includegraphics{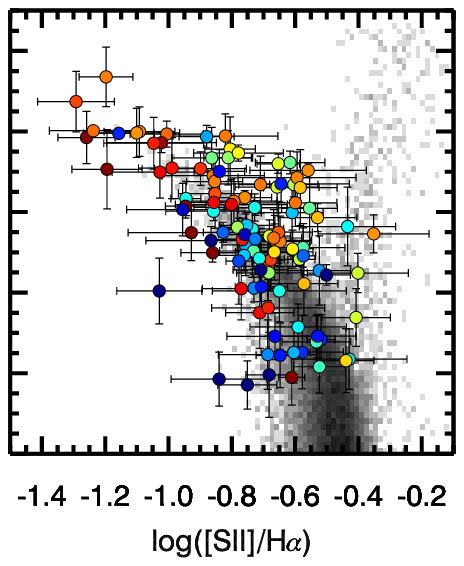}
\hspace{-0.75in}
\includegraphics{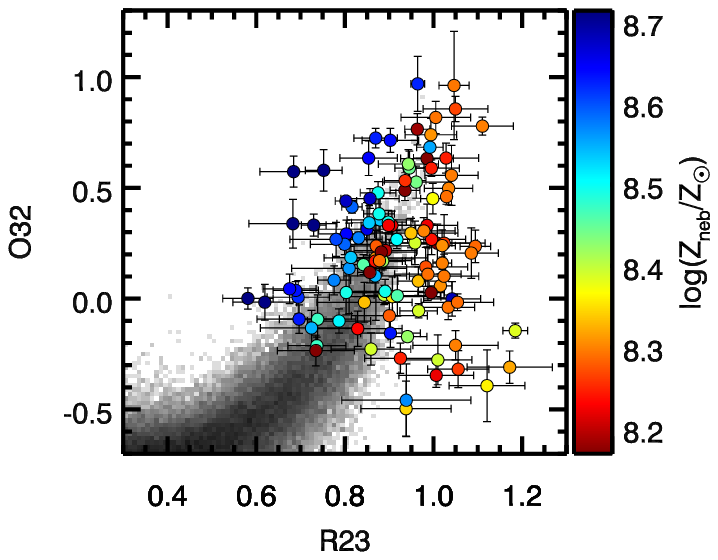}\\
\vspace{-0.7in}
\hspace{-0.1in}
\includegraphics{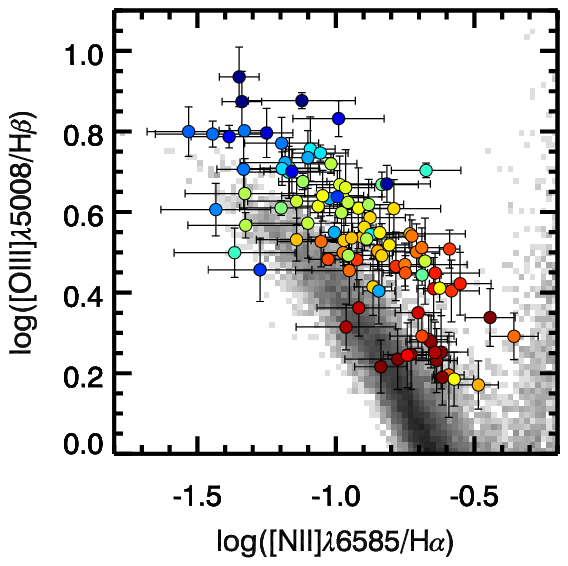}
\hspace{-1.35in}
\includegraphics{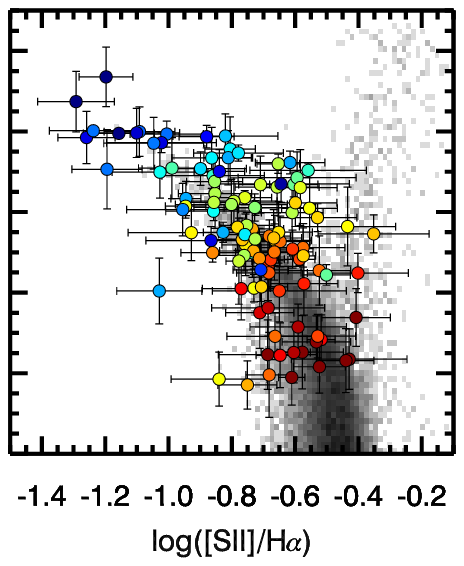}
\hspace{-0.75in}
\includegraphics{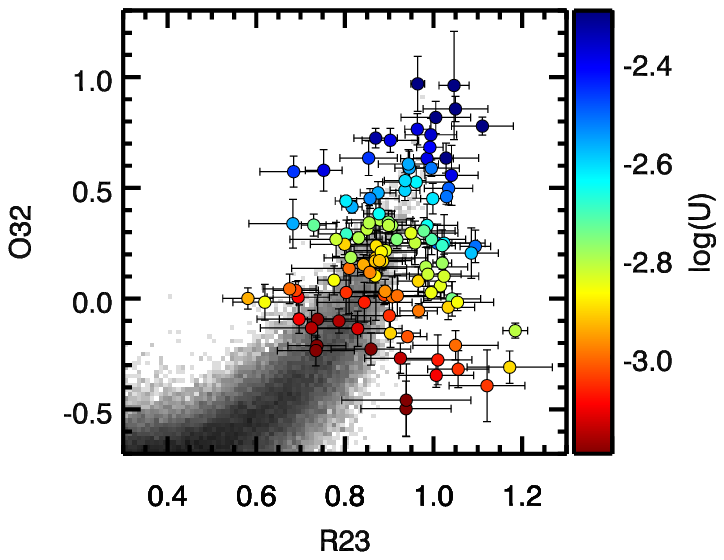}\\
\vspace{-0.7in}
\hspace{-0.1in}
\includegraphics{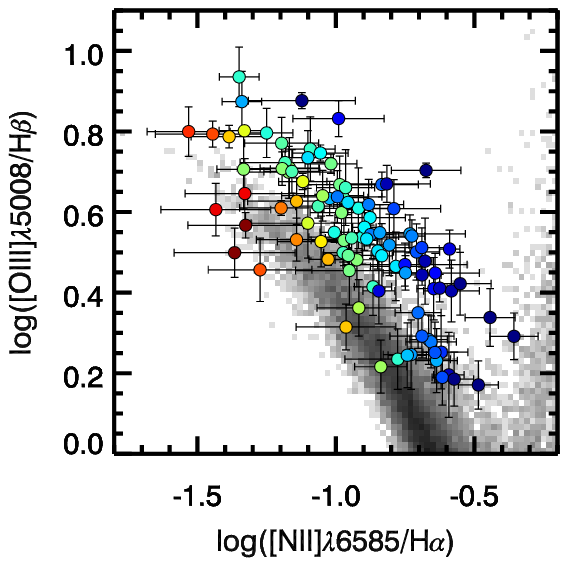}
\hspace{-1.35in}
\includegraphics{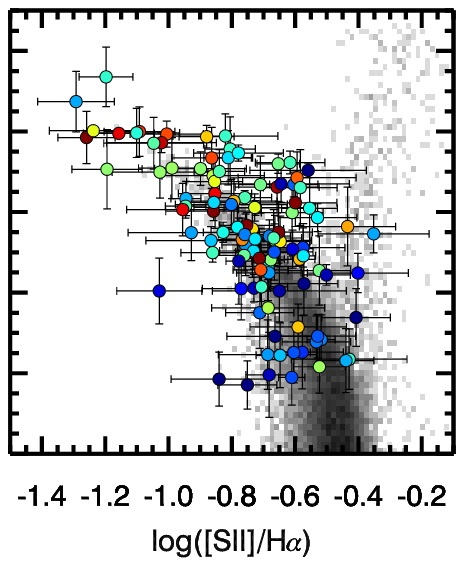}
\hspace{-0.75in}
\includegraphics{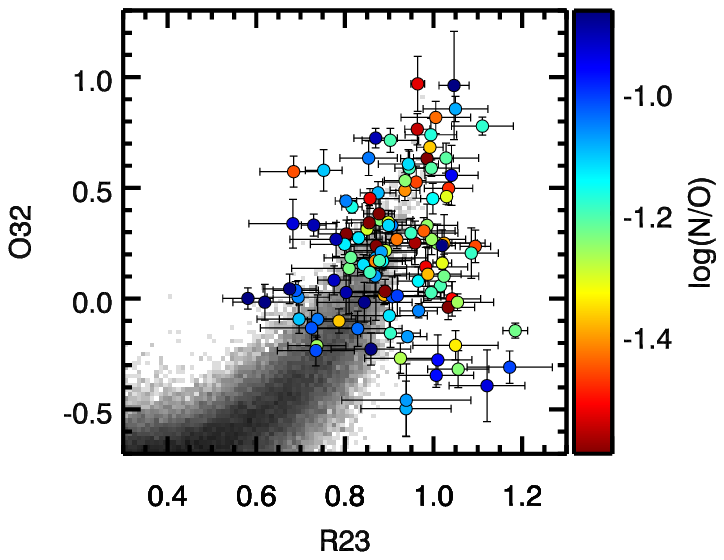}\\
\vspace{-0.7in}
\hspace{-0.1in}
\includegraphics{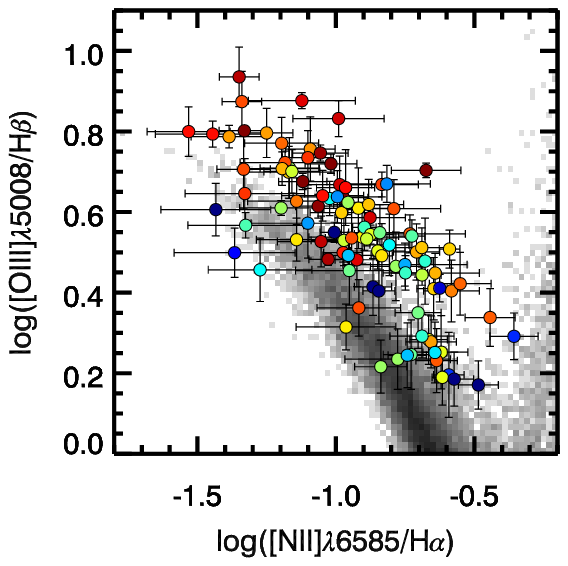}
\hspace{-1.35in}
\includegraphics{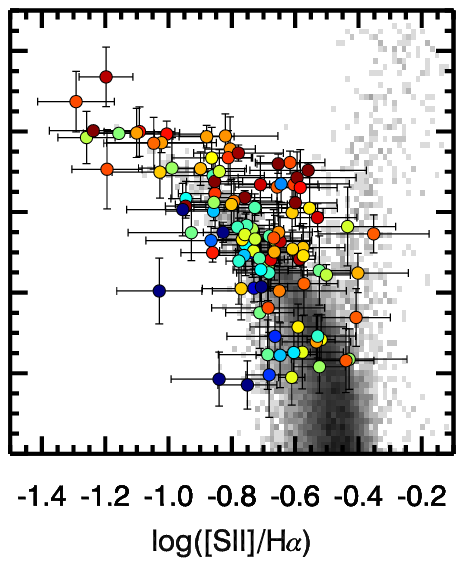}
\hspace{-0.75in}
\includegraphics{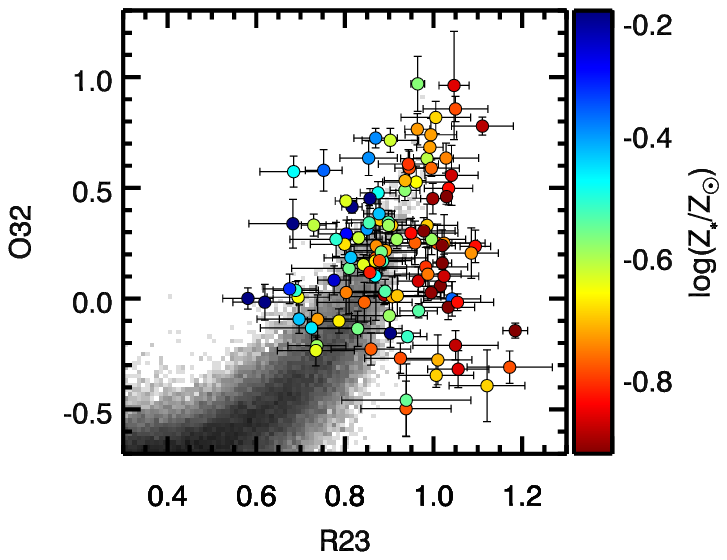}
\caption[Model parameter trends in the classic line-ratio diagrams]{The N2-BPT (left column), S2-BPT (center column), and O32-R23 (right column) diagrams, showing the trends with $\log(Z_{\textrm{neb}}/Z_{\odot}$) (top row), $\log(U)$ (second row), and $\log(\textrm{N/O})$ (third row), and $\log(Z_{\ast})$ (bottom row) for the KBSS sample described in Section~\ref{sample_selection}, with $z\sim0$ SDSS galaxies shown in greyscale for comparison. The KBSS points are color-coded by the inferred value of each parameter and highlight a few key correlations: $Z_{\textrm{neb}}$ and $Z_{\ast}$ vary across the galaxy locus in the O32-R23 diagram (top and bottom right) while showing a weaker dependence on the location in the N2- and S2-BPT diagrams. In contrast, $U$ clearly decreases along the locus in all three diagrams, toward lower [\ion{O}{3}]/H$\beta$ and O32. Inferred N/O increases toward higher [\ion{N}{2}]/H$\alpha$, but also weakly along the locus in the S2-BPT plane, toward lower [\ion{O}{3}]/H$\beta$.}
\label{bpt_trends}
\end{figure*}

It is interesting to consider whether galaxies with certain combinations of physical conditions are more likely to inhabit specific regions of the nebular diagnostic diagrams, resulting in trends in physical conditions along or across the galaxy locus in line-ratio space. For example, the locus of $z\sim0$ star-forming galaxies from SDSS in the N2-BPT diagram is known to be a sequence in both gas-phase O/H (corresponding to $Z_{\textrm{neb}}$) and $U$. However, when we investigate the presence of similar trends among $\langle z\rangle = 2.3$ galaxies (Figure~\ref{bpt_trends}), we find that although $U$ is strongly correlated with position along the galaxy locus in the N2-BPT, the S2-BPT, \textit{and} the O32-R23 diagram (second row of panels), the trend with $Z_{\textrm{neb}}$ in the N2-BPT or S2-BPT planes is noticeably weaker (top left and center panels). The inferred value of $Z_{\textrm{neb}}$ for $\langle z\rangle = 2.3$ galaxies appears to change \textit{across} the galaxy locus rather than \textit{along} it in the O32-R23 diagram, where higher values of $Z_{\textrm{neb}}$ are found on the low-R23 side of the locus. That the trends with $Z_{\textrm{neb}}$ and $U$ are essentially perpendicular to one another in the O32-R23 diagram suggests that a combination of these indices could be used to construct a strong-line diagnostic that simultaneously provides constraints on both O/H and $U$. This idea formed the basis of the earliest strong-line calibrations for R23 \citep[e.g.][]{pagel1979,mcgaugh1991} and has been discussed more recently by \citet{shapley2015}, who showed that the local sequence of galaxies in the O32-R23 diagram increased monotonically in O/H as O32 and R23 declined. We explore the possibility of constructing such a diagnostic using the results for the KBSS sample in Section~\ref{oh_section}.

We also observe a strong correlation between galaxies' positions in the N2-BPT plane and the value of log(N/O) inferred from their nebular spectra (left panel in the third row of Figure~\ref{bpt_trends}), with higher N/O corresponding to higher values of [\ion{N}{2}]/H$\alpha$. This trend is also observed, although much less strongly, in the S2-BPT diagram, with higher N/O occurring in galaxies on the lower part of the locus. This is the opposite of the observed trend with $U$, which decreases in the same direction in the S2-BPT plane. This may suggest the presence of an inverse correlation between the ionization conditions and star-formation history as probed by chemical abundance patterns (at least, N/O) in high-$z$ galaxies, but the orthogonal trend with $Z_{\textrm{neb}}$ in the line ratio diagrams suggests that any such relationship may be qualitatively different from the strong anti-correlation observed between $U$ and O/H at $z\sim0$ \citep[e.g.,][]{dopita1986}. We quantitatively investigate the correlations between inferred physical conditions, including ionization and metallicity, in Section~\ref{correlation_section}.

Finally, in the bottom row of Figure~\ref{bpt_trends}, we investigate the correlation between model-inferred $Z_{\ast}$ and the location of galaxies in nebular parameter space. Consistent with the analysis presented in \citetalias{steidel2016} and \citetalias{strom2017}, the galaxies with the lowest inferred $Z_{\ast}$ (red points)---and, thus the hardest ionizing radiation fields---are among the most offset upward and to the right of the local distribution of galaxies in the N2-BPT and S2-BPT diagrams. Galaxies with smaller [\ion{O}{3}]/H$\beta$ ratios and low R23 at fixed O32 exhibit a more mixed distribution of $Z_{\ast}$, as we might expect from the location of galaxies with multiple peaks in their $Z_{\ast}$ PDFs (Section~\ref{peaks_and_limits}). As with $Z_\textrm{neb}$ (top row of Figure~\ref{bpt_trends}), the overall trend with $Z_{\ast}$ is \textit{across} the high-$z$ galaxy locus; in other words, lines of constant $Z_{\ast}$ appear to be largely parallel to the galaxy locus, especially in the S2-BPT (bottom center panel) and O32-R23 (bottom right panel) diagrams.

\subsubsection{Emission Line SNR}
\label{snr_section}

For the galaxies with bound posteriors, we characterize their marginalized posteriors using four quantities: the MAP estimate for the model parameter, the width of the 68\% HDI (hereafter $\Delta_{68}$), the skewness, and the kurtosis excess. The skewness of a posterior, $S$, quantifies whether the distribution is left- or right-tailed and provides some insight regarding the constraints imposed on a given parameter in certain regions of parameter space. The kurtosis excess, $K$, measures the ``peakiness'' of the distribution (a normal distribution has $K=0$); positive values indicate that the distribution contains more power in the peak relative to the tails, when compared with a normal distribution. As a posterior may have fat tails and still be relatively narrow, we use $\Delta_{68}$ as our primary measure of precision.

Importantly, there appear to be no strong trends in $\Delta_{68}$ for any model parameter in terms of the location of galaxies in the nebular diagnostic diagrams. This result implies that we should be capable of recovering all high-$z$ galaxies' physical conditions with similar precision, given a minimum set of emission line measurements. The exception is galaxies falling below the galaxy locus in the S2-BPT diagram, where objects frequently have parameter PDFs with multiple strong peaks (as shown in Figure~\ref{ratio_diagrams}) and would likely benefit from additional constraints from rest-UV observations.

\begin{figure}
\centering
\includegraphics{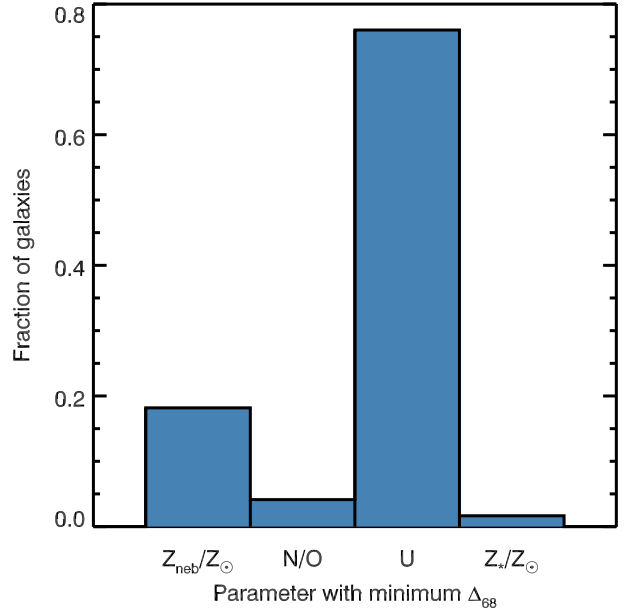}
\caption[Most precisely inferred model parameters for KBSS galaxies]{The distribution of the most precisely-inferred parameter for individual galaxies. In $\sim80$\% of cases, log($U$) is the parameter with the narrowest 68\% HDI ($\Delta_{68}$), suggesting that the observed nebular spectrum responds most sensitively to changes in the ionization conditions and thus has the most power to discriminate between different ionization parameters. In contrast, log($Z_{\ast}/Z_{\odot}$) is the most precise parameter for only two objects, reflecting the relative indirectness of the constraints on the shape of the ionizing radiation obtained from nebular spectroscopy, especially when compared to direct observation of the non-ionizing rest-UV spectum.}
\label{best_cloudy_param}
\end{figure}

Figure~\ref{best_cloudy_param} shows the distribution of the most-precisely inferred parameter (i.e., the parameter with the smallest value of $\Delta_{68}$, corresponding to the fractional error) for individual galaxies. For nearly $\sim80$\% of galaxies, log($U$) is the parameter with the narrowest marginalized posterior, meaning that the nebular spectra of individual galaxies are only consistent with a relatively limited range in $U$, but could be matched by a broader range in $Z_{\textrm{neb}}$, N/O, or $Z_{\ast}$ at a given $U$. In contrast, log($Z_{\ast}/Z_{\odot}$) is the most precise parameter for only two galaxies in the sample, consistent with our argument in Section~\ref{peaks_and_limits} regarding the lack of robust constraints on the shape of the ionizing radiation when using only indirect observables, such as nebular emission lines.

\begin{deluxetable*}{lccccccccccc}
\tablecaption{Correlations between posterior shape and emission line SNR}
\tablehead{
\multicolumn{2}{c}{Posterior quantity} & \multicolumn{2}{c}{Balmer decrement SNR} & \multicolumn{2}{c}{[\ion{O}{3}]$\lambda5008$ SNR} & \multicolumn{2}{c}{[\ion{O}{2}] SNR} & \multicolumn{2}{c}{[\ion{N}{2}]$\lambda6585$ SNR} & \multicolumn{2}{c}{[\ion{S}{2}] SNR} \\
 & & \multicolumn{1}{c}{$\rho$} & \multicolumn{1}{c}{$\sigma$} & \multicolumn{1}{c}{$\rho$} & \multicolumn{1}{c}{$\sigma$} & \multicolumn{1}{c}{$\rho$} & \multicolumn{1}{c}{$\sigma$} & \multicolumn{1}{c}{$\rho$} & \multicolumn{1}{c}{$\sigma$} & \multicolumn{1}{c}{$\rho$} & \multicolumn{1}{c}{$\sigma$}
 }
 \startdata
  log($Z_{\textrm{neb}}/Z_{\odot}$) & $S$ & $+0.17$ & $1.9$ & $+0.05$ & $0.6$ & $+0.09$ & $1.0$ & $+0.02$ & $0.3$ & $-0.10$ & $1.1$ \\
  log($Z_{\textrm{neb}}/Z_{\odot}$) & $K$ & $+0.14$ & $1.5$ & $+0.14$ & $1.6$ & $+0.11$ & $1.2$ & $+0.19$ & $2.1$ & $+0.33$ & $3.5$ \\
  log($Z_{\textrm{neb}}/Z_{\odot}$) & $\Delta_{68}$ & $-0.56$ & $6.1$ & $-0.46$ & $5.0$ & $-0.29$ & $3.1$ & $-0.26$ & $2.8$ & $-0.30$ & $3.3$ \\
\hline
                           log(N/O) & $S$ & $-0.10$ & $1.1$ & $-0.01$ & $0.1$ & $-0.15$ & $1.6$ & $-0.48$ & $5.3$ & $-0.14$ & $1.6$ \\
                           log(N/O) & $K$ & $-0.07$ & $0.8$ & $-0.22$ & $2.4$ & $+0.09$ & $0.9$ & $+0.75$ & $8.2$ & $+0.35$ & $3.8$ \\
                           log(N/O) & $\Delta_{68}$ & $-0.48$ & $5.3$ & $-0.14$ & $1.6$ & $-0.45$ & $4.8$ & $-0.86$ & $9.4$ & $-0.70$ & $7.6$ \\
\hline
                           log($U$) & $S$ & $-0.52$ & $5.7$ & $-0.16$ & $1.8$ & $-0.54$ & $5.7$ & $-0.27$ & $3.0$ & $-0.40$ & $4.3$ \\
                           log($U$) & $K$ & $-0.53$ & $5.8$ & $-0.24$ & $2.7$ & $-0.46$ & $4.9$ & $-0.19$ & $2.1$ & $-0.29$ & $3.1$ \\
                           log($U$) & $\Delta_{68}$ & $-0.62$ & $6.8$ & $-0.49$ & $5.4$ & $-0.39$ & $4.1$ & $-0.37$ & $4.0$ & $-0.36$ & $4.0$ \\
\hline
          log($Z_{\ast}/Z_{\odot}$) & $S$ & $-0.10$ & $1.1$ & $-0.02$ & $0.2$ & $-0.11$ & $1.1$ & $-0.17$ & $1.8$ & $-0.11$ & $1.2$ \\
          log($Z_{\ast}/Z_{\odot}$) & $K$ & $+0.27$ & $3.0$ & $+0.26$ & $2.8$ & $+0.03$ & $0.3$ & $+0.11$ & $1.2$ & $+0.31$ & $3.4$ \\
          log($Z_{\ast}/Z_{\odot}$) & $\Delta_{68}$ & $-0.62$ & $6.8$ & $-0.52$ & $5.7$ & $-0.19$ & $2.1$ & $-0.15$ & $1.6$ & $-0.29$ & $3.1$    
\enddata
\tablecomments{The reported values indicate the Spearman rank correlation coefficient ($\rho$) and the number of standard deviations ($\sigma$) by which the measured sum-squared difference of ranks deviates from the null hypothesis (i.e., no correlation).}
\label{corr_w_observables}
\end{deluxetable*}

Table~\ref{corr_w_observables} lists the Spearman coefficients ($\rho$) and significance of the correlations between $\Delta_{68}$, $S$, and $K$ for each model parameter and the SNR of the Balmer decrement and the most commonly-measured strong emission lines. The skewness and kurtosis excess of the model parameter posteriors (corresponding to the shape of the PDFs) appear most commonly correlated with the SNR of [\ion{S}{2}]$\lambda\lambda6718,6732$, although $S_{\textrm{N/O}}$ and $K_{\textrm{N/O}}$ are also negatively and positively correlated with the SNR of [\ion{N}{2}]$\lambda6585$, respectively; the correlation with $K_{\textrm{N/O}}$ indicates that a higher SNR corresponds to strongly peaked posteriors. In contrast, $S_{U}$ and $K_{U}$ are anti-correlated with the SNR of the Balmer decrement and [\ion{O}{2}]$\lambda3727,3729$ (instead reflecting a tendency toward fat-tailed distributions at high SNR).

Of greater interest is that the precision of the model parameter estimates ($\Delta_{68}$) is sensitive to the quality of specific emission line features. Notably, improvements in $\Delta_{68}$ for log($Z_{\textrm{neb}}/Z_{\odot}$) correlate most strongly with the SNR of the Balmer decrement, but are also significantly (albeit only moderately, i.e., $|\rho|\lesssim0.5$) correlated with all of the other emission features excluding [\ion{N}{2}]$\lambda6585$; this suggests that there is no single feature that is most important for determining $Z_{\textrm{neb}}$ in high-$z$ galaxies, and improving the SNR of any given line measurement only moderately improves the precision of $Z_{\textrm{neb}}$ estimates. Conversely, the precision of inferred log(N/O) is most strongly correlated with the SNR of [\ion{N}{2}]$\lambda6585$, followed by the SNR of [\ion{S}{2}]$\lambda\lambda6718,6732$, the Balmer decrement, and [\ion{O}{2}]$\lambda\lambda3727,3729$. Ionization parameter is the only quantity where $\Delta_{68}$ is significantly correlated with the SNR of all the strongest emission line features, with the precision of the parameter estimate most sensitive to changes in the quality of the Balmer decrement and [\ion{O}{3}]$\lambda5008$ measurements. Somewhat surprisingly, $\Delta_{68}$ for log($Z_{\ast}/Z_{\odot}$) is only significantly correlated with the SNR of the Balmer decrement, [\ion{O}{3}]$\lambda5008$, and [\ion{S}{2}]$\lambda\lambda6718,6732$, perhaps reflecting the greater utility of the S2-BPT diagram over the N2-BPT diagram in discriminating between sources of excitation for high-$z$ galaxies. The advantage of having robust measurements of the [\ion{S}{2}]$\lambda\lambda6718,6732$ doublet is also demonstrated by the fact that only the SNR of the sulfur lines and the Balmer decrement (critical for correctly accounting for differential reddening due to dust) are significantly correlated with the precision of the model estimates for all four parameters.

Collectively, these results have a number of implications for using strong emission line measurements to infer galaxies' physical conditions. First, \textit{the nebular spectrum of high-$z$ galaxies is most sensitive to changes in the ionization conditions in the galaxies' \ion{H}{2} regions}. Consequently, ionization parameter, $U$, is the most easily and precisely determined quantity given the observables that are typically available for high-$z$ galaxies. Although there is certainly some ability to discriminate between the likely shape of the ionizing radiation (parameterized in our model by $Z_{\ast}$), our results also underscore that it is preferable to use more direct constraints, such as observations at rest-UV wavelengths, when available.
 
\section{New Calibrations for Nebular Strong-Line Diagnostics}
\label{strongline_section}

The photoionization model method we have described can easily be applied to the majority of high-$z$ galaxy samples with existing rest-optical spectroscopy. However, it is useful to test whether there are new calibrations for the commonly-used strong-line diagnostics (especially for N/O and O/H) that may be more appropriate for high-$z$ samples than calibrations based on samples at lower redshifts. Significant effort in the last several years has been dedicated to understanding the evolution of strong-line metallicity indices as a function of redshift \citep[e.g.,][among others]{kewley2013,steidel2014,jones2015,shapley2015,cullen2016,dopita2016,sanders2016,hirschmann2017,kashino2017}, with the prevailing wisdom being that calibrations based on ``extreme'' low-$z$ samples of galaxies or \ion{H}{2} regions are better suited for use at $\langle z\rangle = 2.3$ than those based on typical local galaxies. Although what is considered extreme varies, this assumption must be true to some degree, as evidence suggests that typical $\langle z\rangle = 2.3$ galaxies have considerably higher nebular ionization and excitation than typical local galaxies. But it is also true that $\langle z\rangle = 2.3$ galaxies differ in their characteristic star-formation histories relative to galaxies at $z<1$. Such differences will result in important variations in the chemical abundance patterns between high-$z$ galaxies and even ``extreme'' low-$z$ objects, which may, in turn, introduce systematic biases when diagnostics tuned to the latter are applied to the former.

In this section, we investigate the correlation between model-inferred physical conditions for individual galaxies in KBSS and strong-line ratios measured from their nebular spectra. Such an exercise allows us to both comment on which parameters galaxies' spectra are most sensitive to and offer guidance regarding the utility of strong-line diagnostics at high redshift.

\begin{deluxetable*}{llcccc}
\tablecaption{Correlations between strong-line ratios and model-inferred parameters}
\tablehead{
Index & Definition & Parameter & $\rho$\footnotemark[2] & Significance\footnotemark[3] ($\sigma$) & $\sigma_{\textrm{RMS}}$ (dex)}
\startdata
O32\footnotemark[1] & $\log$([\ion{O}{3}]$\lambda\lambda4960,5008$/[\ion{O}{2}]$\lambda\lambda3727,3729$) & $U$ & $+0.91$ & $9.4$ & $0.11$ \\
Ne3O2 & $\log$([\ion{Ne}{3}]$\lambda3869$/[\ion{O}{2}]$\lambda\lambda3727,3729$) & $U$ & $+0.56$ & $4.3$ & $0.22$ \\
O3 & $\log$([\ion{O}{3}]$\lambda5008$/H$\beta$) & $U$ & $+0.83$ & $9.1$ & $0.16$ \\
N2O2\footnotemark[1] & $\log$([\ion{N}{2}]$\lambda6585$/[\ion{O}{2}]$\lambda\lambda3727,3729$) & N/O & $+0.77$ & $7.2$ & $0.12$ \\
N2S2 & $\log$([\ion{N}{2}]$\lambda6585$/[\ion{S}{2}]$\lambda\lambda6718,6732$) & N/O & $+0.82$ & $6.8$ & $0.09$ \\
\vspace{-0.15cm} &  & N/O & $+0.81$ & $8.0$ & $0.11$ \\
\vspace{-0.15cm} N2 &  $\log$([\ion{N}{2}]$\lambda6585$/H$\alpha$) \\
 &  & O/H & +0.31 & $3.0$ & 0.16 \\
O3N2 & $\log$([\ion{O}{3}]$\lambda5008$/H$\beta$)$-\log$([\ion{N}{2}]$\lambda6585$/H$\alpha$) & O/H & $-0.37$ & $3.6$ & $0.16$ \\
R23\footnotemark[1] & $\log$[([\ion{O}{3}]$\lambda\lambda4960,5008$+[\ion{O}{2}]$\lambda\lambda3727,3729$)/H$\beta$] & O/H & $-0.69$ & $6.9$ & $0.09$
\enddata
\footnotetext[1]{Requires an extinction correction; we have adopted the \citet{cardelli1989} extinction curve.}
\footnotetext[2]{The rank correlation coefficient computed using a Spearman test.}
\footnotetext[3]{The number of standard deviations by which the measured correlated deviates from the null hypothesis.}
\label{index_table}
\end{deluxetable*}

\subsection{Ionization Parameter}

\begin{figure*}
\centering
\includegraphics{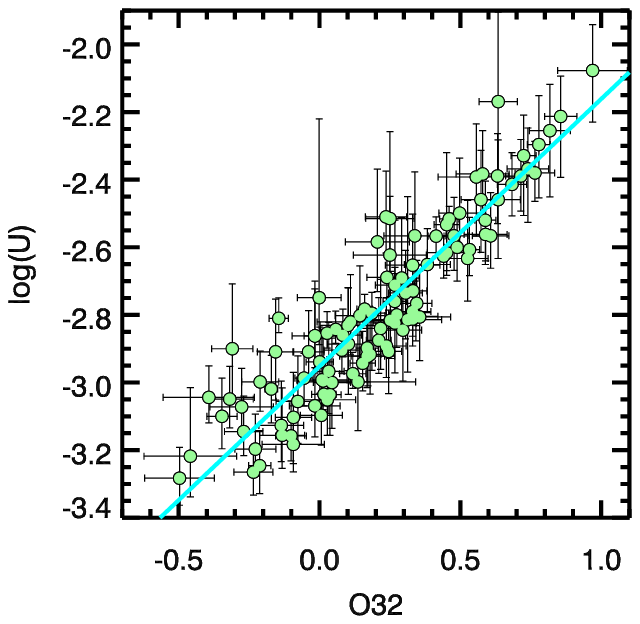}
\hspace{-0.8in}
\includegraphics{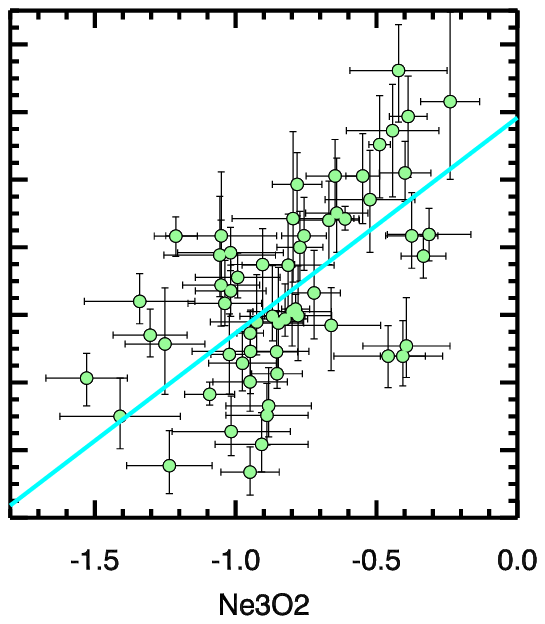}
\hspace{-0.8in}
\includegraphics{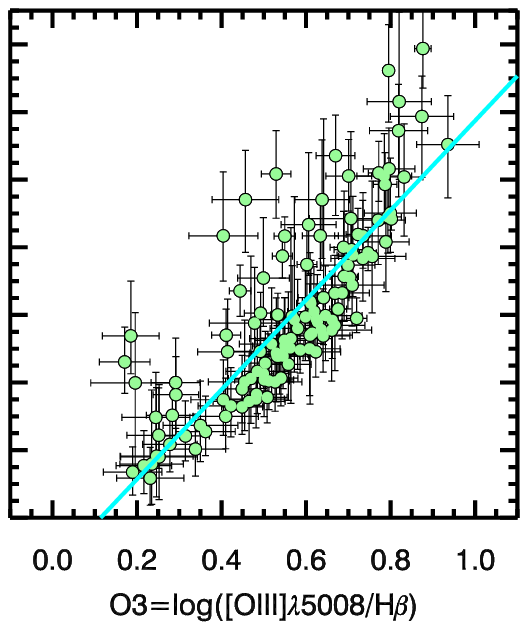}
\caption[Correlation between log($U$) and strong-line indices]{The correlation between model-inferred log($U$) and common strong-line indices: O32, Ne3O2, and O3. The green points show the distribution of KBSS galaxies, limited to those with $\textrm{SNR}>2$ for the line indices used. Unsurprisingly, O32 shows the strongest correlation and least scatter with log($U$). Although not commonly used to infer log($U$), O3 is also strongly correlated with ionization parameter. Ne3O2 exhibits large scatter, which can in part be attributed to the lower typical SNR for [\ion{Ne}{3}]$\lambda3869$ measurements than for [\ion{O}{3}]$\lambda5008$ or [\ion{O}{2}]$\lambda\lambda3727,3729$. The cyan lines show the best-fit linear relations determined using the KBSS results from our photoionization model method.}
\label{logu_calibs}
\end{figure*}

The importance of the ionization state of the gas in \ion{H}{2} regions in determining the resulting nebular spectrum is obvious: in order for the majority of the emission line features to be observed, H must be photoionized and then recombine; simultaneously, heavier elements such as N, O, and S must be singly- or doubly-ionized, then collisionally excited. The details of the observed collisionally-excited emission are set by $T_e$, which is in turn sensitive to the distribution of ionizing photon energies and the gas cooling (dominated by emission from O atoms). Oxygen is the only element with multiple ions whose commonly-observed transitions fall in the $J$-, $H$-, and $K$-band for galaxies with $z\simeq2-2.7$\footnote{Doubly-ionized S has a low-lying transition at 9533\AA, but this is only observable from the ground for galaxies with $z\lesssim1.5$.}, making the combination of [\ion{O}{3}]$\lambda5008$ and [\ion{O}{2}]$\lambda\lambda3727,3729$ measurements a sensitive probe of the ionization conditions in high-$z$ galaxies. Although [\ion{O}{1}]$\lambda6301$ can be also be observed in the spectra of individual $z\simeq2-2.7$ galaxies on occasion, emission from neutral O may not be spatially coincident with emission from ionized O, which requires at least an H-ionizing photon to be created. We discuss the challenges associated with observations of ions that trace low-ionization gas in Section~\ref{no_section} below.

As discussed in Section~\ref{grid_section}, we parameterize the ionization state of gas using the ionization parameter, $U=n_{\gamma}/n_H$. Higher values of $U$ will result in more doubly-ionized O relative to neutral and singly-ionized O in the irradiated gas. In general, $n_{\gamma}$ can be varied by changing the shape and/or the normalization of the ionizing radiation field, so measurements of $U$ are most meaningful when details about the ionizing source are also known. In our method, we explicitly decouple these effects so that $U$ reflects the required scaling of the ionizing radiation field, while the shape is set by the $Z_{\ast}$ of the input BPASSv2 model.

Figure~\ref{logu_calibs} shows the correlation between model-inferred log($U$) and common strong-line indices for the sample of KBSS galaxies with bound posteriors in all model parameters and $\textrm{SNR}>2$ for the line index (different for each panel). The definitions for O32, Ne3O2, and O3 are listed in Table~\ref{index_table}, along with other commonly-used strong-line indices. Although all three indices shown in Figure~\ref{logu_calibs} are strongly positively correlated with $U$, the correlation with O32 is the strongest and most significant (Spearman $\rho=0.91$, significance of 9.4$\sigma$). We can determine a calibration based on the KBSS results \citep[calculated using the \textit{MPFITEXY} IDL routine\footnote{\url{http://purl.org/mike/mpfitexy}},][]{williams2010}, which has the following form:
\begin{equation}
\log(U) = 0.79\times\textrm{O32}-2.95.
\label{logu_equation}
\end{equation}
Relative to the best-fit relation (shown in cyan), the measurements have RMS scatter $\sigma_{\textrm{RMS}} = 0.11$~dex. The same statistics for the correlations between log($U$) and both Ne3O2 and O3 are reported in Table~\ref{index_table}, corresponding to the following calibrations:
\begin{eqnarray}
&&\log(U) = 0.64\times\textrm{Ne3O2}-2.22 \\
&&\log(U) = 1.33\times\textrm{O3}-3.55.
\end{eqnarray}

\begin{figure*}
\centering
\includegraphics{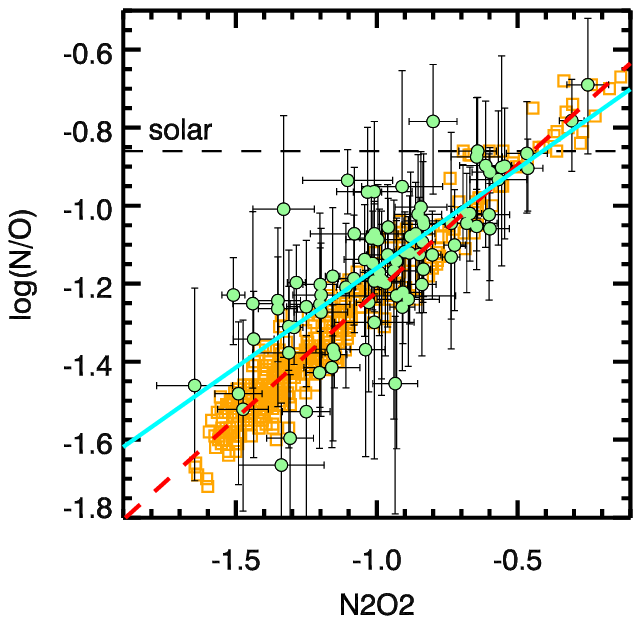}
\hspace{-0.8in}
\includegraphics{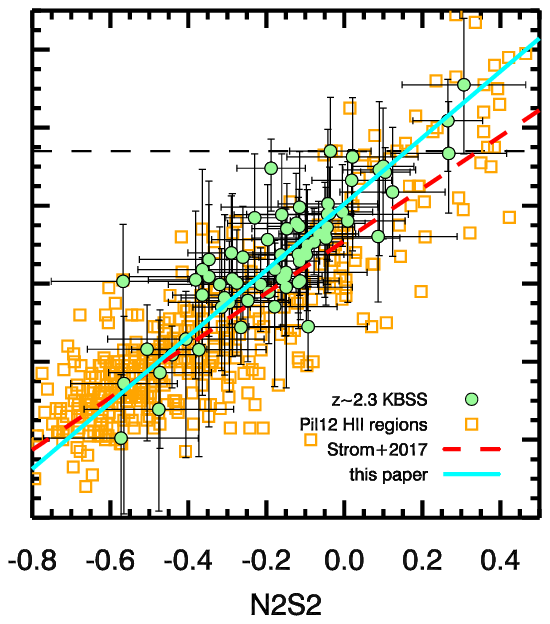}
\hspace{-0.8in}
\includegraphics{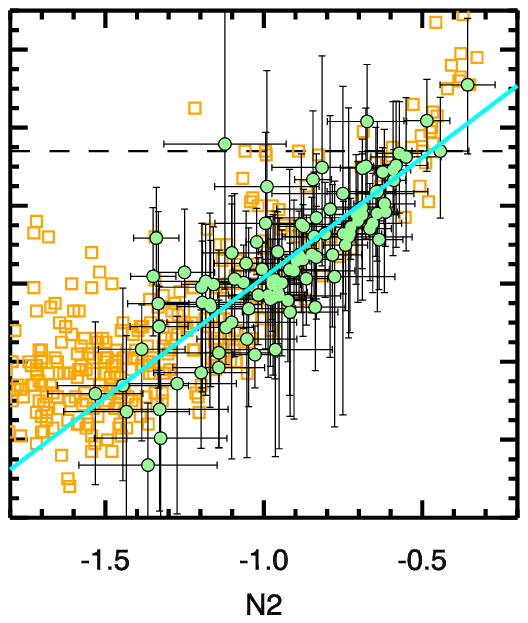}
\caption[Correlation between log(N/O) and strong-line indices]{The correlation between model-inferred log(N/O) and common strong-line indices: N2O2, N2S2, and N2. As in Figure~\ref{logu_calibs}, the green points show the location of KBSS galaxies with $\textrm{SNR}>2$ for the line indices used. For comparison, local \ion{H}{2} regions from \citetalias{pilyugin2012} are shown as orange squares. The black dashed horizontal line shows the solar value from \citet{asplund2009}. The red dashed lines show the calibrations determined using a sample of local \ion{H}{2} regions from \citetalias{pilyugin2012}, whereas the solid cyan lines show the calibrations determined from the KBSS sample.}
\label{no_calibs}
\end{figure*}

It is perhaps somewhat disappointing that the correlation between Ne3O2 and log($U$) is so poor, given the practical advantages of measuring Ne3O2 relative to O32. Because [\ion{Ne}{3}]$\lambda3869$ and [\ion{O}{2}]$\lambda\lambda3727,3729$ are close in wavelength, observing both emission features requires less observing time for high-$z$ galaxies, and uncertainties in dust reddening are reduced. However, despite the relatively tight locus in Ne3O2 vs. O32 space observed for high-ionization $z\sim0$ SDSS galaxies \citep[c.f.][]{levesque2014}, $z\sim2$ galaxies show notably more scatter in the same observed line ratio space \citepalias{strom2017}. We also showed in \citetalias{strom2017} that changes in the shape of the ionizing radiation field at fixed $U$ affect Ne3O2 more strongly than O32. For these reasons, the larger intrinsic scatter between Ne3O2 and log($U$) may not be entirely unexpected.

In contrast, it is interesting that O3 (which has the same advantages as Ne3O2) is more significantly correlated with log($U$) and has less scatter about the best-fit calibration. An important caveat is that the calibrations presented here are only appropriate for objects with stellar populations similar to those of KBSS galaxies; this is especially important for O3, which is more strongly correlated with O/H in samples with presumably softer ionizing radiation fields, including SDSS galaxies \citep{maiolino2008}.


\subsection{Nitrogen-to-oxygen Ratio}
\label{no_section}

In contrast with $U$, N/O has a smaller effect on the overall nebular spectra from \ion{H}{2} regions, as it primarily affects emission lines of N. Still, as with $U$, there are clear and direct proxies for measuring N/O accessible in the rest-optical nebular spectra of galaxies. Due to similarities in the ionization potentials of N and O, the ionization correction factors are also relatively similar, meaning that N$^+$/O$^+$ corresponds roughly to N/O. As [\ion{N}{2}]$\lambda6585$ and [\ion{O}{2}]$\lambda\lambda3727,3729$ are frequently detected in high-$z$ galaxy spectra, N/O is one of the most accessible probes of the chemical abundance pattern in galaxies' interstellar medium (ISM).

The topic of N/O in galaxies has a long history, with a recent resurgence in interest as samples of high-$z$ galaxies with rest-optical spectra have increased in size and quality \citep[e.g.,][]{masters2014,shapley2015,masters2016,kashino2017,kojima2017}. In \citetalias{strom2017}, we addressed the issue of N/O in KBSS galaxies, using N2O2 and a calibration based on a sample of extragalactic \ion{H}{2} regions from \citetalias{pilyugin2012}. The 414 objects from \citetalias{pilyugin2012} have direct method measurements of N/H and O/H, which make them a useful comparison sample that we also choose to employ here.

Figure~\ref{no_calibs} shows the correlation between model-inferred log(N/O) and N2O2, N2S2 (another commonly-used probe of N/O), and N2 for KBSS galaxies (green points) and the \citetalias{pilyugin2012} \ion{H}{2} regions (orange squares). Also shown are calibrations based on the \citetalias{pilyugin2012} sample from \citetalias{strom2017} (red dashed lines) and new calibrations determined based on the work presented in this paper (cyan lines). As with the strong-line calibrations for $U$ introduced above, we assess the usefulness of the strong-line ratios as proxies for N/O by calculating the strength and significance of the correlations, as well as the best-fit linear relation between N/O and the indices. The results of this analysis are listed in Table~\ref{index_table}, and the calibrations themselves are
\begin{eqnarray}
&&\log(\textrm{N/O}) = 0.51\times\textrm{N2O2}-0.65 \label{n2o2_equation} \\
&&\log(\textrm{N/O}) = 0.85\times\textrm{N2S2}-1.00\\
&&\log(\textrm{N/O}) = 0.62\times\textrm{N2}-0.57.
\end{eqnarray}
All three indices have similarly significant correlations with N/O, as determined by a Spearman rank-correlation test, and similar RMS scatter relative to the new calibrations. We can take some guidance from the distribution of \citetalias{pilyugin2012} \ion{H}{2} regions, which exhibit similar behavior in the indices relative to N/O, but have smaller measurement errors. For the \citetalias{pilyugin2012} sample, N2O2 (which is the most direct proxy for N/O) exhibits the smallest scatter relative to the best-fit relation based on the same sample (the red dashed line in the left panel of Figure~\ref{no_calibs}); this suggests that N2O2 may also be the most accurate probe of N/O in high-$z$ galaxies.

\begin{figure}
\centering
\includegraphics{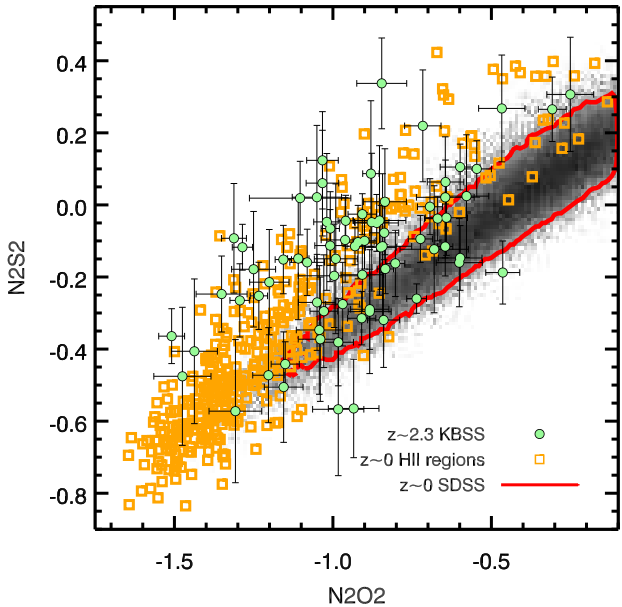}
\caption[Comparison between N2S2 and N2O2 indices]{N2S2 and N2O2 for three samples of objects: $\langle z\rangle = 2.3$ KBSS galaxies (green points), individual $z\sim0$ \ion{H}{2} regions from \citetalias{pilyugin2012} (orange squares), and $z\sim0$ galaxies from SDSS (greyscale, with 90\% of the sample enclosed by the red contour). As in Figure~\ref{no_calibs}, KBSS galaxies and the \citetalias{pilyugin2012} sample exhibit similar behavior, suggesting that N2O2 and N2S2 trace N/O in similar ways between the two samples. However, the locus of SDSS galaxies occupies a distinct region of parameter space, with somewhat higher N2O2 and N2S2 overall, but lower N2S2 at a given N2O2 than either the $\langle z\rangle = 2.3$ galaxies or $z\sim0$ \ion{H}{2} regions. One explanation for this difference may be the increased importance of low-ionization diffuse ionized gas in the integrated-light spectra of nearby galaxies.}
\label{n2s2_vs_n2o2}
\end{figure}

Although the calibrations based on the \citetalias{pilyugin2012} sample are similar to those based on our new measurements for individual $\langle z\rangle = 2.3$ KBSS galaxies (especially for N2O2), we note that calibrations based on samples of local \textit{galaxies} generally fare much worse in returning accurate estimates of N/O for high-$z$ galaxies. Figure~\ref{n2s2_vs_n2o2} shows the reason for this discrepancy. Although N2S2 and N2O2 have almost identical behavior relative to one another for KBSS galaxies (green points) and \citetalias{pilyugin2012} \ion{H}{2} regions (orange squares), the behavior of the same indices for $z\sim0$ galaxies from SDSS (greyscale, with 90\% of galaxies enclosed by the red contour) differs significantly, particularly at higher values of N2S2 and N2O2. Just as locally-calibrated strong-line methods for O/H that use the N2-BPT lines will return inconsistent estimates relative to one another for objects that fall significantly off the $z\sim0$ N2-BPT locus, strong-line methods for N/O based on $z\sim0$ galaxies will disagree for high-$z$ galaxies.

This inconsistency poses challenges to easily comparing N/O inferred using N2S2 and N2O2. \citet{masters2016} argue that the N/O-M$_{\ast}$ relation evolves slowly with redshift, with high-$z$ galaxies exhibiting only slightly lower N/O at fixed M$_{\ast}$ relative to local galaxies. Their inferences are based on results from \citet{kashino2017}, who examine the behavior of N2S2 with M$_{\ast}$ for a sample of $z\sim1.6$ galaxies from the FMOS-COSMOS program \citep{silverman2015} relative to $z\sim0$ galaxies from SDSS and find that the values of N2S2 observed in high-mass galaxies in their sample approach the N2S2 measured in high-mass SDSS galaxies. Based on these observations and the expectation that the mass-metallicity relation evolves more quickly with redshift \citep[e.g.,][]{erb2006metal,steidel2014,sanders2015}, \citet{masters2016} claim that the slight observed decrement in N2S2 of the $z\sim1.6$ FMOS-COSMOS galaxies relative to $z\sim0$ SDSS galaxies reflects higher N/O values in high-$z$ galaxies at fixed O/H relative to the N/O-O/H relation observed in the local universe. From our analysis of N/O in KBSS galaxies in \citetalias{strom2017}, however, we show that the evolution in the N/O-M$_{\ast}$ relation is roughly equivalent in magnitude to the inferred evolution in the O/H mass-metallicity relation, with $\langle z\rangle = 2.3$ galaxies exhibiting values of log(N/O) that are $\sim0.32$~dex lower than $z\sim0$ galaxies at fixed M$_{\ast}$. While we could not directly study the N/O-O/H relation for high-$z$ galaxies in \citetalias{strom2017}, we revisit the likelihood of elevated N/O at fixed O/H in Section~\ref{no_offset_section}.

The behavior of N2S2 and N2O2 shown in Figure~\ref{n2s2_vs_n2o2} offers some clues that may explain the discrepancy between our interpretation of the data and the interpretation favored by \citet{masters2016} and \citet{kashino2017}. We can confidently assume that N/O increases with increasing N2S2 and N2O2 for all three samples. Nevertheless, N2S2 does not increase as quickly with N/O in $z\sim0$ galaxies as it does in $z\sim0$ \ion{H}{2} regions and $\langle z\rangle = 2.3$ galaxies. As a result, a comparison between N2S2, without first converting the index to N/O using the appropriate calibration, will underestimate N/O in $z\sim0$ galaxies relative to $\langle z\rangle = 2.3$ galaxies. Conversely, comparing N2O2 between $z\sim0$ and $\langle z\rangle = 2.3$ galaxies (as we did in \citetalias{strom2017} by assuming the same calibration for N/O) may over-estimate the difference between the two samples, but N2O2 remains a far more direct tracer of N/O than N2S2.

The comparison between N2S2 and N2O2 shown in Figure~\ref{n2s2_vs_n2o2} also highlights an intriguing and likely very meaningful result: $\langle z\rangle = 2.3$ galaxies have nebular spectra that are much more similar to individual \ion{H}{2} regions at $z\sim0$ than to intregrated-light spectra of $z\sim0$ galaxies. There are a number of reasons this might be the case, but as other authors \citep[including][]{sanders2016,kashino2017} have also suggested, the most straightforward explanation is that star-formation in $\langle z\rangle = 2.3$ galaxies takes place mostly in one, or a few, dominant \ion{H}{2} regions, compared to typical $z\sim0$ galaxies with similar M$_{\ast}$, where \ion{H}{2} regions are more isolated with respect to one another. Thus, an integrated-light spectrum of a local galaxy may potentially include contributions from gas outside \ion{H}{2} regions (especially for neutral species like \ion{O}{1} and low-ionization species like S$^+$, which do not require an H-ionizing photon to be created); we refer readers to \citet{sanders2017} for more information regarding the biases introduced by not accounting for this diffuse ionized gas. In this context, the behavior of N2S2 for SDSS galaxies in Figure~\ref{n2s2_vs_n2o2} makes more sense. If the [\ion{S}{2}]$\lambda\lambda6718,6732$ emission observed in SDSS galaxy spectra includes contributions from the diffuse ionized gas in addition to \ion{H}{2} region emission, N2S2 will be depressed at fixed N2O2 relative to high-$z$ galaxies and individual \ion{H}{2} regions.

\subsection{Oxygen Abundance}
\label{oh_section}

\begin{figure*}
\centering
\includegraphics{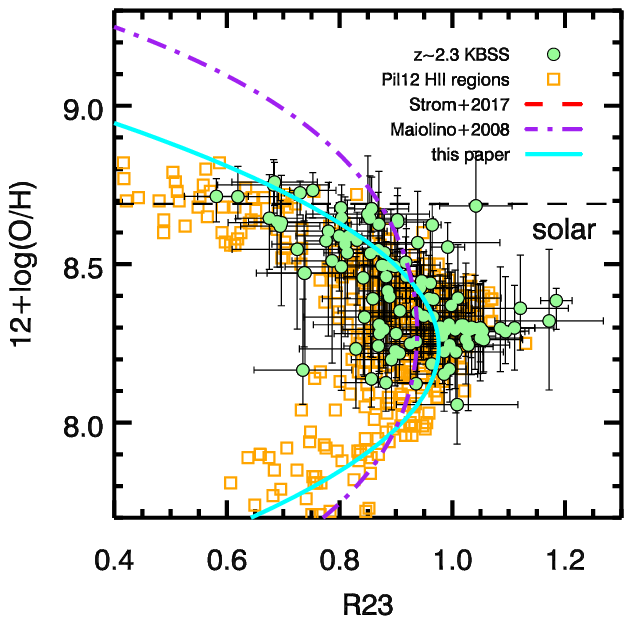}
\hspace{-0.8in}
\includegraphics{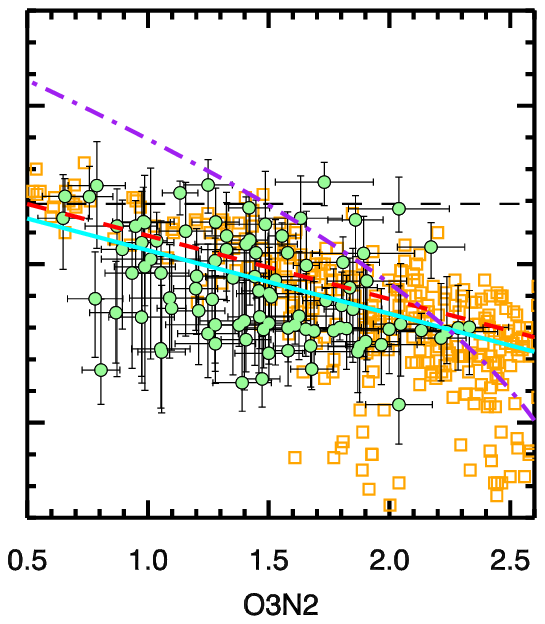}
\hspace{-0.8in}
\includegraphics{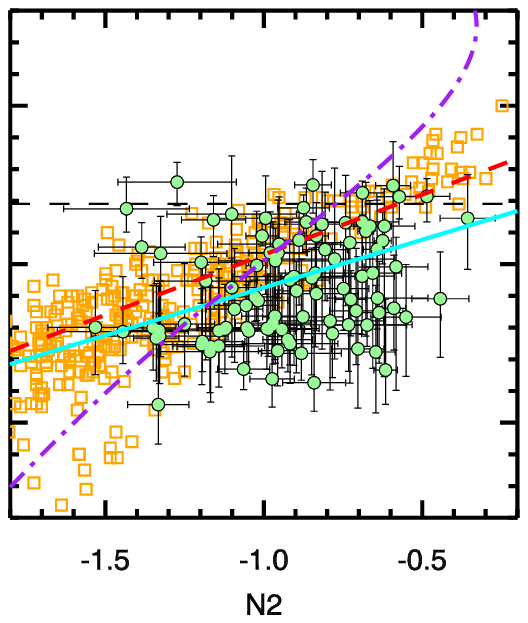}
\caption[Correlation between $12+\log(\textrm{O/H})$ and strong-line indices]{The correlation between $12+\log(\textrm{O/H})$ and the most commonly-used strong-line diagnostics for O/H, shown in the same manner as Figure~\ref{no_calibs}. The $\langle z\rangle = 2.3$ KBSS points are illustrated as green points with error bars, using the value of their model-inferred O/H. The $z\sim0$ \citetalias{pilyugin2012} sample is shown as orange squares, using O/H estimated using the direct method and shifted toward higher O/H by $0.24$~dex to account for the difference in abundance scale, as described in the text. The red dashed lines show the calibrations based on the \citetalias{pilyugin2012} sample from \citetalias{strom2017}, and the purple dot-dashed lines represent the calibrations from \citet{maiolino2008}. All three indices are significantly correlated with O/H for KBSS galaxies, although R23 shows markedly less scatter about the best-fit relations (shown in cyan).}
\label{oh_calibs}
\end{figure*}

Of perhaps greatest interest is the potential to provide new calibrations for strong-line diagnostics for O/H, which is generally the quantity one intends when referring to gas-phase ``metallicity''. The impact of the overall metallicity of ionized gas on the resulting nebular spectrum is more nuanced than the effects of ionization and differences in abundance \textit{ratios}, like N/O. Metals, like O, are the primary coolants in low-density ionized gas, as they are able to convert kinetic energy in the gas into electromagnetic emission through collisionally-excited transitions. Yet, trends in emission line strength with O/H for transitions such as [\ion{O}{3}]$\lambda5008$ are frequently complicated due to the competition between enrichment and gas cooling. The classic example is the double-valued behavior of R23, which increases with increasing O/H due to the larger number of O atoms present---until some critical value (usually near $12+\log(\textrm{O/H})=8.3$, but sensitive to the shape of the ionizing radiation field), at which point gas cooling becomes so efficient that the collisional excitation of O$^{+}$ and O$^{++}$ drops, and the observed value of R23 declines. Galaxies or \ion{H}{2} regions with maximal values of R23($\approx0.8-1.0$) must therefore have moderate O/H, as lower or higher values of O/H would result in a lower value of R23 being observed. This reasoning led us to conclude in \citetalias{steidel2016} and \citetalias{strom2017} that most $\langle z\rangle = 2.3$ KBSS galaxies must be moderately O-rich, which we have now confirmed in this paper (Figure~\ref{cloudy_hists}).

For strong-line indices other than R23, including O3N2 and N2, part of the ability to estimate O/H comes from indirectly measuring N/O instead (by including [\ion{N}{2}]$\lambda6585$) and relying on the existence of a relationship between N/O and O/H. This implicit dependence on the N/O-O/H relation is what, until now, has prevented a direct investigation of the N/O-O/H relation at high redshift (the subject of Section~\ref{no_vs_oh_section}) and an independent analysis of O/H enrichment in the early universe, free from the biases introduced by relying on local calibrations.

We are now poised to develop new calibrations for O/H using the results from our photoionization model method, but must first address the issue of abundance scales. To this point, we have reported the results of our photoionization model method in terms of $Z_{\textrm{neb}}$. However, given the importance of O to the overall mass budget of metals in \ion{H}{2} regions and, thus, to gas cooling and the nebular spectrum, it is reasonable to assume that $Z_{\textrm{neb}}$ largely traces gas-phase O/H. As a result, we may convert log($Z_{\textrm{neb}}/Z_{\odot}$) to the more commonly-used  metric $12+\log(\textrm{O/H})$ by adopting a fiducial solar value. From \citet{asplund2009}, $12+\log(\textrm{O/H})_{\odot}=8.69$, so $12+\log(\textrm{O/H})_{\textrm{KBSS}}=\log(Z_{\textrm{neb}}/Z_{\odot})+8.69$.

The O/H scale that results from this translation differs significantly from the abundance scale for diagnostics determined using $T_e$-based measurements of O/H, which are lower by $\approx0.24$~dex relative to measurements based on nebular recombination lines (see the discussion in Section~8.1.2 of \citealt{steidel2016}; also \citealt{esteban2004,blanc2015}). Authors who have investigated this phenomenon report that the abundances resulting from recombination line methods are in fact closer to the \textit{stellar} abundance scale than abundances from $T_e$-based methods that rely on collisionally-excited emission lines. Additionally, as we showed in \citetalias{strom2017}, a similar offset in 12+log(O/H) is required to force (N/O)$_{\odot}$ to correspond with 12+log(O/H)$_{\odot}$ for local \ion{H}{2} regions from \citetalias{pilyugin2012}, where both N/O and O/H are measured using the direct method. Therefore, in order to compare our O/H measurements with those for the \ion{H}{2} region sample used in the previous section (which is also the same as used in \citetalias{strom2017}), we shift the \citetalias{pilyugin2012} abundances toward higher $12+\log(\textrm{O/H})$ by 0.24~dex. It is not necessary to account for this difference when considering the abundance \textit{ratio} N/O, as it affects inferences of both O/H and N/H.

Figure~\ref{oh_calibs} compares the correlation between $12+\log(\textrm{O/H})$ and the three most commonly-used strong-line indices for measuring O/H for both $\langle z\rangle = 2.3$ KBSS galaxies (green points) and the local \ion{H}{2} regions from \citetalias{pilyugin2012} (orange squares). Notably, R23 shows the strongest trend with $12+\log(\textrm{O/H})$ for $\langle z\rangle = 2.3$ KBSS galaxies, with O3N2 and N2 showing markedly larger scatter. Table~\ref{index_table} lists the Spearman coefficient and the significance of the correlation for all three indices.

As with strong-line indices and N/O in Figure~\ref{no_calibs}, the behavior of R23 with O/H appears very similar to that observed for the \citetalias{pilyugin2012} \ion{H}{2} regions, although the KBSS sample mostly populates the high-metallicity ``branch.'' If we solve for a quadratic calibration that describes the relation between $12+\log(\textrm{O/H})$ and R23 for the KBSS sample (shown in cyan in the left panel of Figure~\ref{oh_calibs}), we find
\begin{equation}
12+\log(\textrm{O/H}) = 8.24+(0.85-0.87\times\textrm{R23})^{1/2},
\label{r23_equation}
\end{equation}
with the R23 measurements exhibiting $\sigma_{\textrm{RMS}} = 0.09$~dex relative to the calibration.

The calibration in Equation~\ref{r23_equation} is significantly different from the calibration for R23 provided by \citet{maiolino2008}, which we have shifted by 0.24~dex and shown as the dot-dashed purple curve in the left panel of Figure~\ref{oh_calibs}. \citet{maiolino2008} used a superset of direct method and photoionization model measurements of O/H to construct a family of self-consistent strong-line diagnostics (the corresponding calibrations for O3N2 and N2 are shown in the center and right panels). Although both the \citet{maiolino2008} and KBSS calibrations reach a maximum value of R23 at nearly the same oxygen abundance ($12+\log(\textrm{O/H})\approx 8.3$) the KBSS calibration has a smaller latus rectum\footnote{The latus rectum of a conic section, here a parabola, is the line segment that passes through the focus and is perpendicular to the major axis. The length of the latus rectum describes the ``width" of the parabola.}, with R23 varying more quickly as a function of O/H. We note, however, that $\sim84$\% of $\langle z\rangle = 2.3$ KBSS galaxies have $\textrm{R23}>0.8$, where the index (regardless of calibration choice) becomes relatively insensitive to changes in O/H. As a result, even our re-calibration for R23 is primarily useful for O-rich galaxies.

\begin{figure}
\centering
\includegraphics{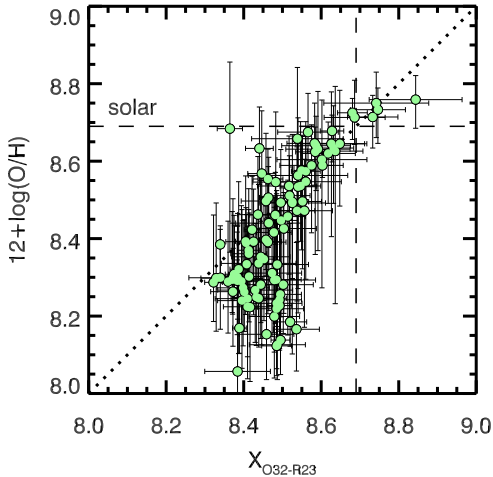}
\caption[Combined O/H calibration using O32 and R23]{$12+\log(\textrm{O/H})$ as a function of $X_{\textrm{O32-R23}}$, a diagnostic combining O32 and R23, as described in the text. The dotted line shows the one-to-one relation. Although making use of both indices reduces scatter relative to the model-inferred O/H at high-metallicities, the calibration fails at $12+\log(\textrm{O/H})\lesssim8.6$ ($Z_{\textrm{neb}}/Z_{\odot}\lesssim0.8$).}
\label{o32r23_calib}
\end{figure}

If we attempt to construct a calibration based on O32 \textit{and} R23, in order to jointly account for the effects of O/H and $U$, we can reduce the scatter relative to the model-inferred O/H at high metallicities, as shown in Figure~\ref{o32r23_calib}. Here, we have chosen a diagnostic with the following form:
\begin{equation}
\begin{split}
12+\log(\textrm{O/H}) = X_{\textrm{O32-R23}} = 9.59+0.93\times\textrm{O32}\\-1.85\times\textrm{R23}+0.18\times\textrm{O32}^2+0.65\times\textrm{R23}^2\\-0.96\times(\textrm{O32}\times\textrm{R23}),
\end{split}
\end{equation}
which is based on the sample of KBSS galaxies occupying the upper branch of the R23 relation, corresponding to $12+\log(\textrm{O/H})\geq8.3$, and determined using a least-squares fit. For large values of O/H, this calibration has less RMS scatter than the R23-only calibration---but it fails for galaxies with $12+\log(\textrm{O/H})\lesssim8.6$ ($Z_{\textrm{neb}}/Z_{\odot}\lesssim0.8$), greatly reducing its utility for samples at high redshift. It is likely that a separate O32-R23 calibration could be determined for galaxies on the low-metallicity branch in R23, but it is clear that inferring O/H for galaxies with R23 near the maximum using a simple strong-line diagnostic is more challenging.

The large scatter between both O3N2 and N2 and O/H for KBSS galaxies is notable and may caution against relying on either index as a diagnostic for O/H in high-$z$ galaxies, despite the formal significance of the correlations. Still, these indices have been favored in the past due to their practical advantages: they are less sensitive to dust reddening and require fewer emission line measurements, thus necessitating less observing time. Using the KBSS results to construct new calibrations, we find
\begin{eqnarray}
&&12+\log(\textrm{O/H}) = 8.75-0.21\times\textrm{O3N2} \\
&&12+\log(\textrm{O/H}) = 8.77+0.34\times\textrm{N2}.
\end{eqnarray}
We emphasize that, even using these calibrations, the large scatter could lead to incorrect estimates of O/H in high-$z$ galaxies. However, the situation is worse if calibrations based on a local sample (the red dashed line in the right panel of Figure~\ref{oh_calibs}) are used instead; our results imply that $12+\log(\textrm{O/H})$ could be overestimated by up to $\sim0.4$~dex for some high-$z$ galaxies.

\begin{figure*}
\centering
\includegraphics[width=0.3\textwidth]{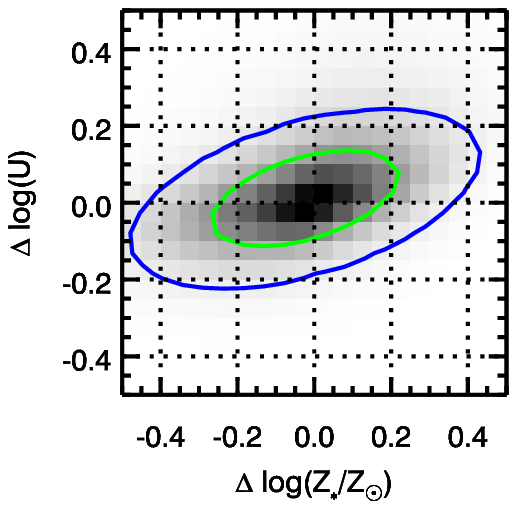}
\includegraphics[width=0.3\textwidth]{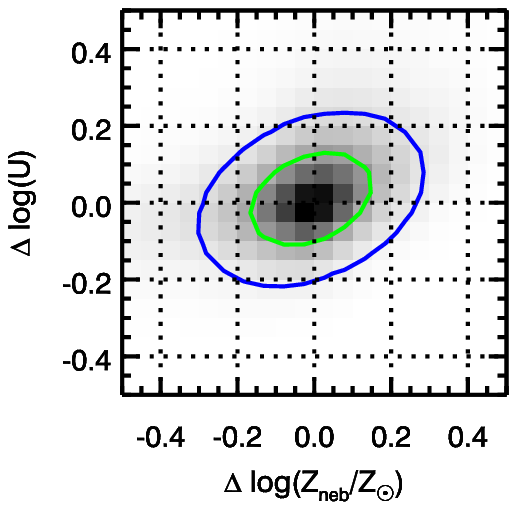}
\includegraphics[width=0.3\textwidth]{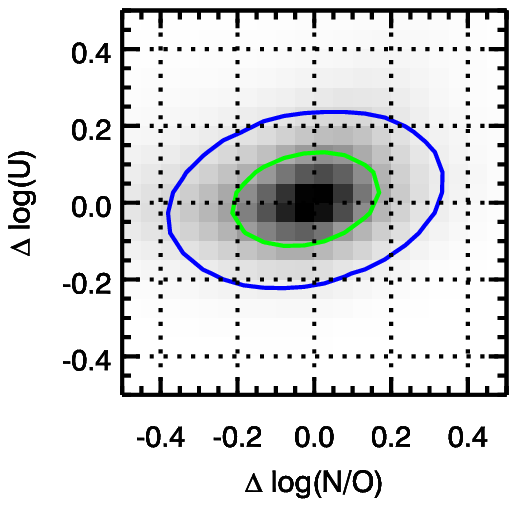}
\vspace{-0.2in}
\\
\includegraphics[width=0.3\textwidth]{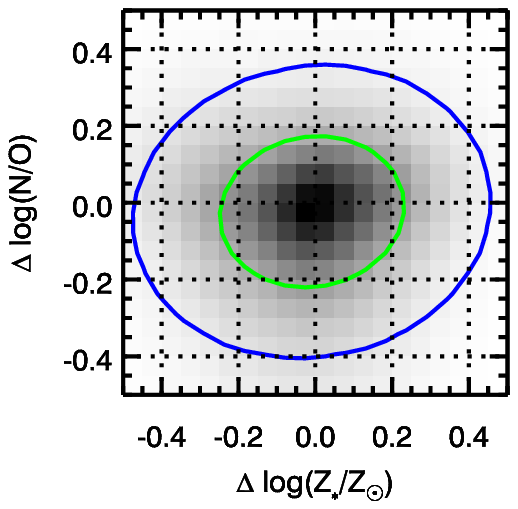}
\includegraphics[width=0.3\textwidth]{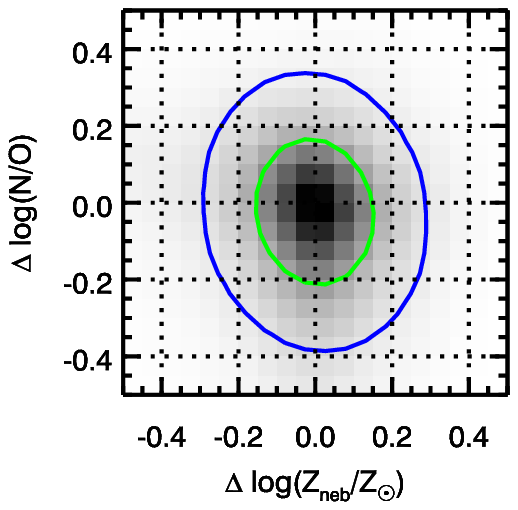}
\includegraphics[width=0.3\textwidth]{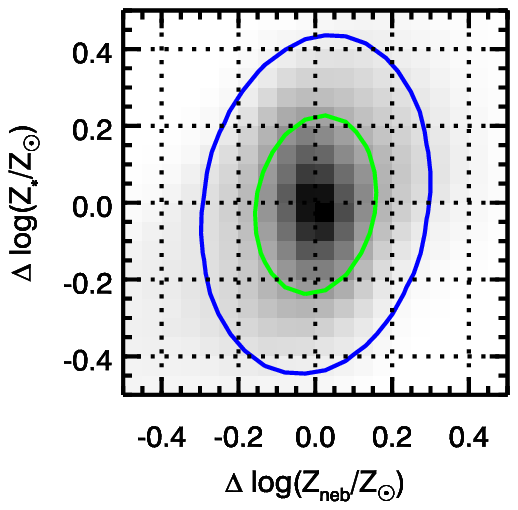}
\caption[Average 2D posterior distributions for model-inferred parameters]{The average 2D posterior distributions for log($Z_{\textrm{neb}}/Z_{\odot}$), log(N/O), log($U$), and log($Z_{\ast}/Z_{\odot}$). The density maps are constructed by shifting the normalized 2D posteriors such that the MAP estimates for the parameters are at the origin and summing the posteriors for all objects with bound posteriors. The blue and green isodensity contours are drawn at 10\% and 50\% of the peak density, assuming the posteriors can be described by a bivariate Gaussian. The 2D posteriors illustrate the typical correlation between parameters as well as the relative precision of the parameter estimates. For example, log($U$) is generally better constrained than log($Z_{\ast}/Z_{\odot}$), but the parameters are highly correlated. In contrast, estimates of log(N/O) and log($Z_{\ast}/Z_{\odot}$) exhibit less correlation with one another, on average.}
\label{param_correlations}
\end{figure*}

Given these results, it would seem that the most accurate and precise estimates of $12+\log(\textrm{O/H})$ for $\langle z\rangle = 2.3$ galaxies \textit{currently} achievable result from using a photoionization model method such as the one described in this paper. Although R23 (or a combined diagnostic using O32 and R23) is useful for galaxies with relatively high metallicities ($12+\log(\textrm{O/H})\gtrsim8.5-8.6$), there is significant added value in using the same emission line measurements to simultaneously constrain O/H, N/O, and $U$. Measurements of $T_e$-based abundances for a representative sample of high-$z$ galaxies, which will be enabled in the future by facilities like the \textit{James Webb Space Telescope}, may lead to improved calibrations. However, we caution that strong-line calibrations for quantities like N/O and O/H inevitably marginalize over intrinsic scatter in the galaxy population---which may itself be correlated with other physical conditions.

\section{Correlations Between Physical Conditions}
\label{correlation_section}

\subsection{Correlations Between Model Parameters for Individual Galaxies}

We have previously noted the similarity in the behavior of the nebular spectra of $\langle z\rangle = 2.3$ KBSS galaxies and individual local \ion{H}{2} regions, especially as compared to $z\sim0$ galaxies. This agreement---or disagreement, in some cases---arises from similarities and differences in the underlying correlations between the physical conditions driving the observed spectrum. Thus, it is important to quantify the intrinsic correlations between these parameters to understand the ways in which the high-$z$ galaxy populations differ from galaxies found in the present-day universe.

When performing linear regression of astronomical data, the measurement errors on the independent and dependent variables are often assumed to be independent. In the case that the measurement errors are actually correlated, however, the magnitude of the observed correlation will be overestimated if the measurement error correlation has the same sign as the intrinsic correlation between variables and underestimated in cases where the measurement error correlation has a different sign from the intrinsic correlation \citep{kelly2007}. As a result, weak correlations where the measurement error correlation has the opposite sign may be unrecoverable unless the measurement errors \textit{and} the covariance between measurement errors have been accounted for using a statistical model.

From the MCMC results shown in Figures~\ref{good_mcmc} and \ref{bad_mcmc}, it is clear that the model parameter estimates are correlated with one another for at least some fraction of the KBSS sample. Knowledge of these degeneracies for individual galaxies assists in measuring the intrinsic correlations between parameters across the entire sample, but it is also interesting to consider the \textit{typical} correlation between model parameters for a single galaxy, which can inform our intuition regarding galaxies' nebular spectra. 

\begin{deluxetable}{llc}
\tablecaption{Typical correlations between model parameters for individual galaxies}
\tablehead{
Parameter 1 & Parameter 2 & \multicolumn{1}{c}{Correlation coeff.}
}
\startdata
log($Z_{\ast}/Z_{\odot}$) & log(N/O) & $+0.07$ \\
log($Z_{\textrm{neb}}/Z_{\odot}$) & log(N/O) & $-0.09$ \\
log($Z_{\textrm{neb}}/Z_{\odot}$) & log($Z_{\ast}/Z_{\odot}$) & $+0.14$ \\
log(N/O) & log($U$) & $+0.21$ \\
log($Z_{\textrm{neb}}/Z_{\odot}$) & log($U$) & $+0.31$ \\
log($Z_{\ast}/Z_{\odot}$) & log($U$) & $+0.48$
\enddata
\tablecomments{The linear correlation coefficient is determined from the covariance matrix derived from the 2D Gaussian fit to the average pairwise posteriors.}
\label{covar_table}
\end{deluxetable}

Figure~\ref{param_correlations} shows the average pairwise posteriors for the 4 model parameters, which have been constructed by shifting the normalized 2D posteriors for individual objects so that the peak of the distribution is located at the origin and summing the PDFs for all objects with bound posteriors. The green and blue contours highlight the shape of the posterior, assuming it can be described by a bivariate Gaussian function, and the linear correlation coefficients derived from the fits are reported in Table~\ref{covar_table}. The correlations between some model parameters are expected, including the strong correlation between $U$ and $Z_{\ast}/Z_{\odot}$ (upper left panel), which parameterize the normalization and shape of the ionizing radiation, respectively. These results highlight the power of applying additional constraints, where available. For example, if we could impose a prior on $Z_{\ast}$ from observations of the rest-UV spectra of the same galaxies (as we did for a composite spectrum in \citetalias{steidel2016}), our estimates of log($U$) would be even more precise because we would not need to marginalize over as large a range in $Z_{\ast}$.

In some cases, the correlations between inferred parameters run counter to our expectations for the intrinsic correlations between the same parameters: N/O and $Z_{\textrm{neb}}/Z_{\odot}$ (bottom center panel) are anti-correlated for a typical individual galaxy, whereas N/O and O/H (which is traced by $Z_{\textrm{neb}}/Z_{\odot}$) are positively correlated for galaxy and \ion{H}{2} region samples in the nearby universe. This result underscores the importance of accounting for the correlations between inferred parameters for individual objects when quantifying the intrinsic correlations between the same parameters in the full sample, which is the focus of the remainder of this section.

\subsection{N/O-O/H Relation} 
\label{no_vs_oh_section}

\begin{figure}
\centering
\includegraphics{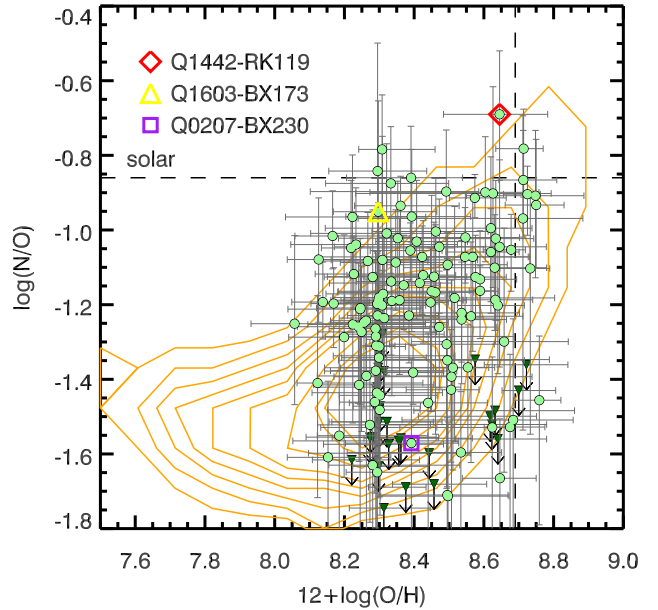}
\caption[The $z\sim2.3$ N/O-O/H relation]{The N/O-O/H relation observed for $\langle z\rangle = 2.3$ KBSS galaxies, shown as green points; galaxies with only a limit on the inferred value of log(N/O) are shown as dark green triangles at the location corresponding to a $2\sigma$ upper limit. For comparison, the distribution of $z\sim0$ objects from \citetalias{pilyugin2012} is represented by the orange contours, with the outermost contour enclosing 99\% of the sample. Most of the KBSS galaxies ($\sim72$\%) fall within these contours and appear to follow the same trend as the local \ion{H}{2} regions. Three galaxies with different combinations of N/O and O/H are highlighted by colored symbols: Q1442-RK119 (red diamond), Q1603-BX173 (yellow triangle), and Q0207-BX230 (purple square). For comparison, the locations of these galaxies in common line-ratio diagrams are shown in Figure~\ref{bpt_no_examples}.}
\label{no_vs_oh_plot}
\end{figure}

Figure~\ref{no_vs_oh_plot} shows that $\langle z\rangle = 2.3$ star-forming galaxies exhibit N/O ratios at fixed O/H consistent with the range observed in local \ion{H}{2} regions. The KBSS sample is shown as green points or dark green triangles (for galaxies with only upper limits on N/O), with the local sample from \citetalias{pilyugin2012} represented by the orange contours; 72\% of KBSS galaxies fall within the outermost contour, which contains 99\% of the \citetalias{pilyugin2012} objects, and there is good agreement in the distribution of N/O at fixed O/H between the samples. We note that the paucity of objects with $12+\log(\textrm{O/H})<8.2$ is likely the result of selection bias. Galaxies with lower M$_{\ast}$ than typical KBSS galaxies, such as the Lyman-$\alpha$-selected galaxies from \citet{trainor2015} with M$_{\ast} \approx 10^{8}-10^{9}$~M$_{\odot}$, are expected to populate this region of parameter space.

Like the $z\sim0$ \ion{H}{2} regions, $\langle z\rangle = 2.3$ galaxies show large scatter in N/O at fixed O/H. At $12+\log(\textrm{O/H})=8.3$ ($Z_{\textrm{neb}}/Z_{\odot}\approx0.4$), the full range in observed $\log(\textrm{N/O})$ is $\sim0.8$~dex, consistent with that observed among the \citetalias{pilyugin2012} \ion{H}{2} regions. Although N/O and O/H are formally significantly correlated for the local sample, a simple Spearman test indicates only a marginally significant (2.3$\sigma$) correlation between the two parameters for KBSS galaxies. Given the anti-correlation between inferred N/O and O/H for individual galaxies (bottom center panel of Figure~\ref{param_correlations}, Table~\ref{covar_table}), however, a Spearman test is likely to underestimate the presence of a positive correlation between N/O and O/H for the KBSS sample. To address this issue, we use the IDL routine LINMIX\_ERR to measure the strength of the correlation between N/O and O/H among the KBSS galaxy sample while accounting for both measurement errors and the covariance between measurement errors. The resulting linear correlation coefficient inferred for the $\langle z\rangle = 2.3$ N/O-O/H relation is $\rho=0.52$, with an overall $99.8$\% likelihood of a positive correlation between the parameters and an estimated intrinsic scatter of $\sigma_{\textrm{int}} = 0.18$~dex.

\begin{figure*}
\centering
\includegraphics{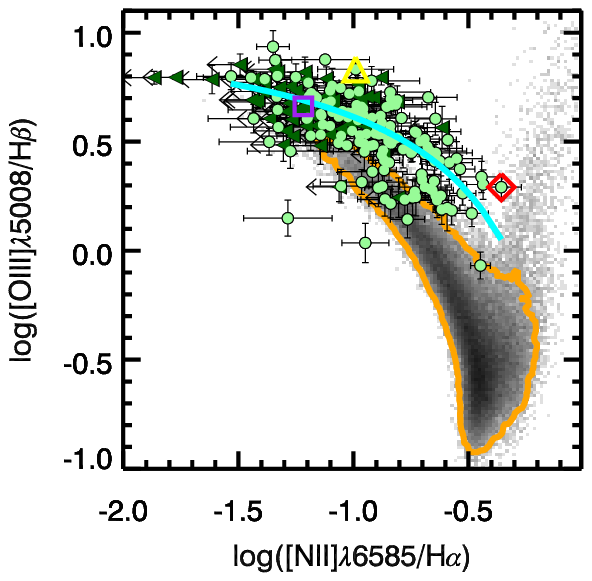}
\hspace{-0.8in}
\includegraphics{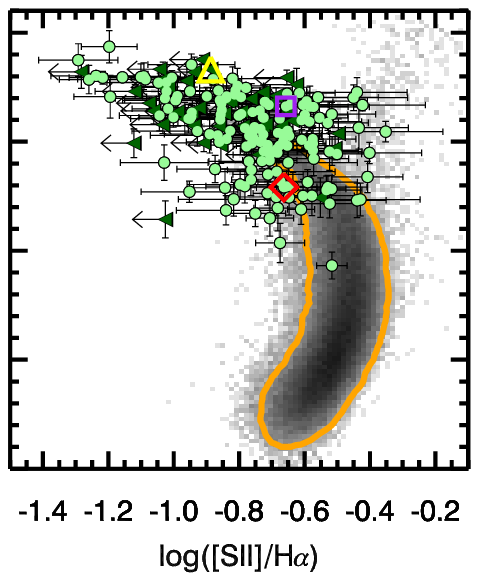}
\includegraphics{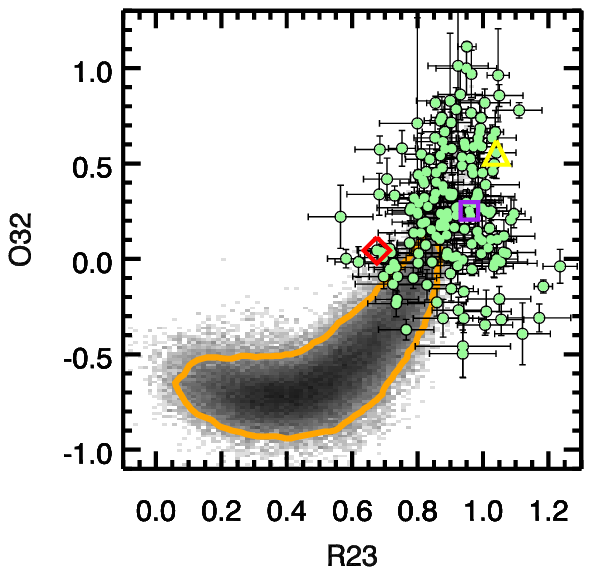}
\caption[Nebular spectra of KBSS galaxies with different N/O and O/H]{Three classic nebular diagnostic diagrams (the N2-BPT, S2-BPT, and O32-R23 diagrams), showing the location of the galaxies highlighted in Figure~\ref{no_vs_oh_plot} in observed line-ratio space; the cyan curve in the left panel reflects the ridge-line of the $z\sim2$ KBSS sample from \citetalias{strom2017}. Q1442-RK119 (identified by the red diamond) has the highest N/O in the sample, and also shows the largest [\ion{N}{2}]/H$\alpha$ ratio. Q1603-BX173 (yellow triangle) and Q0207-BX230 (purple square) have different inferred N/O but very similar O/H, but show significant differences not only in [\ion{N}{2}]/H$\alpha$, but in all three line ratio diagrams. These results demonstrate the challenges associated with attempts to derive physical information from a single 2D projection of multidimensional line-ratio space.}
\label{bpt_no_examples}
\end{figure*}

The physical cause of this intrinsic scatter in N/O at fixed O/H is not immediately obvious for either the $\langle z\rangle = 2.3$ sample or the $z\sim0$ objects. Some authors have found that local galaxies with higher N/O at a given O/H also have higher SFR \citep{andrews2013}, but we find no strong evidence of a similar trend for the KBSS sample. To illustrate differences in the nebular spectra of high-$z$ galaxies found in different regions of the N/O-O/H plane, we highlight three galaxies in Figure~\ref{no_vs_oh_plot}: Q1442-RK119, which has the highest value of N/O in the KBSS sample (red diamond); Q1603-BX173, near the upper envelope of the locus at more moderate O/H (yellow triangle); and Q0207-BX230, which has similar O/H to Q1603-BX173, but is near the low-N/O edge of the distribution (purple square). The locations of these galaxies in the N2-BPT, S2-BPT, and O32-R23 diagrams are also noted in Figure~\ref{bpt_no_examples}. As expected for a galaxy with high N/O, Q1442-RK119 also has a very high value of N2 (the highest among the star-forming KBSS galaxies discussed in this paper\footnote{AGN at similar redshifts can have higher N2 values, which we reported for KBSS in \citetalias{strom2017}. \citet{coil2015} discusses the nebular properties of a more diverse sample of AGN from the MOSDEF survey, which extend to even larger N2.}). Further, in contrast to the two galaxies with more moderate O/H, Q1442-RK119 is found on the low-R23 side of the galaxy locus in the O32-R23 diagram, as expected.

A comparison between the locations of Q1603-BX173 and Q0207-BX230 demonstrates that an increase in N/O at fixed O/H does not correspond only to a horizontal shift in the N2-BPT diagram as one might naively expect; instead, Q1603-BX173 also exhibits higher [\ion{O}{3}]/H$\beta$ and O32 and lower [\ion{S}{2}]/H$\alpha$. Together, these differences reflect a increase in log($U$) of 0.43~dex (at similar O/H to Q0207-BX230), which is clear when information from all three diagrams is incorporated (as in our photoionization model method) but more difficult to diagnose on the basis of the galaxies' locations in a single diagram.

\subsubsection{Enhanced N/O and the N2-BPT Offset}
\label{no_offset_section}

In \citetalias{strom2017}, we studied the distribution of N/O (inferred using N2O2) within the KBSS sample without the aid of independent estimates of O/H. Based on our comparison between $\langle z\rangle = 2.3$ KBSS galaxies and $z\sim0$ galaxies that likely share similar ionizing radiation fields, we concluded that although high-$z$ galaxies may have somewhat higher N/O than local galaxies with the same O/H, this could not account for the entirety of the offset observed between typical $\langle z\rangle = 2.3$ galaxies and typical $z\sim0$ galaxies in the N2-BPT diagram. Now, with self-consistent estimates of N/O and O/H in hand for a large sample of $\langle z\rangle = 2.3$ galaxies, we can revisit the question of whether N/O is enhanced in the most offset star-forming galaxies at high redshift. 

If we simultaneously solve for the best-fit linear relation between N/O and O/H for KBSS galaxies that fall above the sample ridge-line in the N2-BPT diagram (the cyan curve in the left panel of Figure~\ref{bpt_no_examples}) and those that fall below the ridge-line, we confirm an offset between the two samples of $\Delta\log(\textrm{N/O})=0.25\pm0.05$~dex, with the most offset KBSS galaxies exhibiting significantly higher N/O at fixed O/H than KBSS galaxies that appear more similar to typical galaxies in SDSS. However, we note that this difference cannot account for the \textit{total} offset between the KBSS subsamples in the N2-BPT diagram \citepalias[$\approx0.34$~dex,][]{strom2017}, reaffirming our conclusion from \citet{steidel2014} and \citetalias{strom2017} that high-$z$ galaxies must also have harder ionizing radiation fields at fixed O/H relative to typical local galaxies.

Using the results from our MCMC method, we find that a combination of higher ionization parameters and harder spectra (due to more Fe-poor stellar populations) account for the remainder of the offset that cannot be explained by the modest N/O enhancement at fixed O/H. A two-sample KS test reveals that the distributions of $U$ and $Z_{\ast}/Z_{\odot}$ for KBSS galaxies above and below the ridge-line in the N2-BPT diagram are statistically inconsistent with one another. For galaxies above the ridge-line, the median $\log(U)=-2.69$ and the median $Z_{\ast}/Z_{\odot}=0.18$, whereas for galaxies below the ridge-line, the median $\log(U)=-2.90$ and the median $Z_{\ast}/Z_{\odot}=0.25$.

\subsection{$U$-O/H and $U$-$Z_{\ast}$ Relations}

\begin{figure}
\centering
\includegraphics{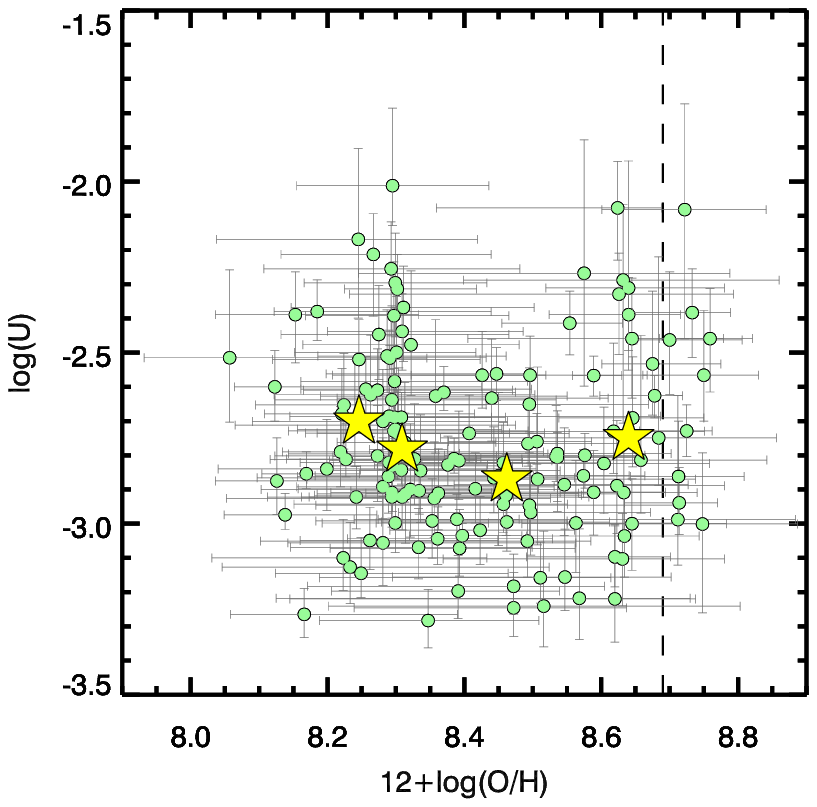}
\caption[The $z\sim2.3$ $U$-O/H relation]{Ionization parameter as a function of gas-phase O/H for 148 $\langle z\rangle = 2.3$ KBSS galaxies. The yellow stars represent the median value of log($U$) in equal-number bins of $12+\log(\textrm{O/H})$. The dashed vertical line represents the solar value of $12+\log(\textrm{O/H})_{\odot}=8.69$. There is no obvious relationship between the two model-inferred parameters for the high-$z$ sample. Even after accounting for measurement errors and covariance between measurement errors, the probability of an anti-correlation is only 29\%.}
\label{logu_vs_oh_plot}
\includegraphics{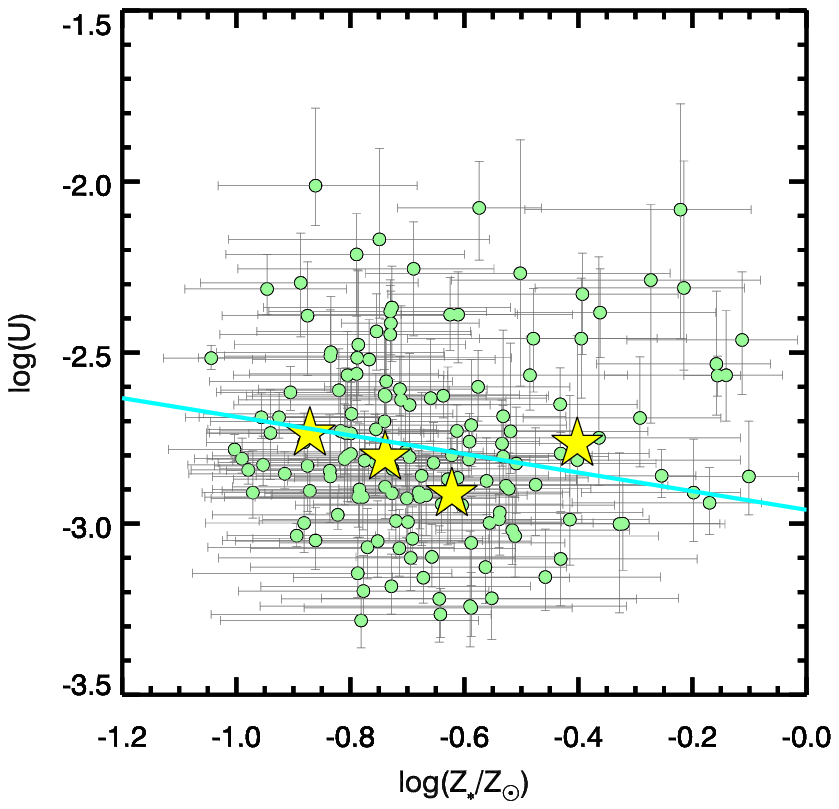}
\caption[The $z\sim2.3$ $U$-$Z_{\ast}$ relation]{Ionization parameter as a function of stellar metallicity, which traces Fe/H, for the same galaxies shown in Figure~\ref{logu_vs_oh_plot}; here, the yellow stars show the median value of log($U$) in equal-number bins of $Z_{\ast}/Z_{\odot}$. Accounting for measurement errors reveals evidence of a somewhat more significant anti-correlation than observed between $U$ and O/H. We calculate a 93\% probability of a negative correlation, with a linear correlation coefficient of $\rho=-0.20$. The best-fit linear relation corresponding to this correlation is shown as a cyan line in the figure.}
\label{logu_vs_zstar_plot}
\end{figure}

Several authors have found evidence for an anti-correlation between ionization parameter and O/H in low-$z$ samples \citep[e.g.,][]{dopita1986,dopita2006,perez-montero2014}, but evidence for a similar trend among high-$z$ galaxies is less abundant. \citet{sanders2016} offer a thorough exploration of the behavior of O32 with galaxy properties in $\langle z\rangle = 2.3$ MOSDEF galaxies; they noted the strong observed anti-correlation between O32 and M$_{\ast}$ and suggested that an anti-correlation between ionization parameter and metallicity at high redshift might also exist, presuming the existence of a mass-metallicity relation. \citet{onodera2016} estimate $q$ ($= Uc$) and O/H for a sample of $z\gtrsim3$ galaxies using strong-line methods and recover a strong anti-correlation, but note that it arises largely because both parameters are estimated using the same line ratio. Although \citet{kojima2017} use $T_e$-based estimates of O/H instead of relying on strong-line methods, they find that most of the $z\sim2$ galaxies in their sample are not consistent with the observed $U$-O/H relation at $z\sim0$.

Figure~\ref{logu_vs_oh_plot} shows little evidence for a correlation between $U$ and O/H in the KBSS sample, even after accounting for measurement errors and the correlation between the inferred parameters for individual galaxies (top center panel of Figure~\ref{param_correlations}). Our results show that there is only a 29\% probability of an intrinsic anti-correlation between $U$ and O/H in our sample, consistent with no correlation. Given our sample selection, however, the range in O/H observed in KBSS galaxies does not extend to very low metallicities, which could impede our ability to recover a weak, but significant anti-correlation in that regime. It is also possible that our definition of $U$, which is a free parameter that does not incorporate differences in geometry or the shape of the ionizing radiation field, obscures an intrinsic relationship between the total ionizing flux and O/H in the population---which has been found at $z\sim0$, often using somewhat different definitions for ionization parameter.

Although unexpected in the context of the local universe \citep[c.f. the steeper relations reported by, e.g.,][]{perez-montero2014,sanchez2015}, the absence of a strong anti-correlation between $U$ and O/H, coupled with the sensitivity of high-$z$ galaxies' nebular spectra to differences in $U$ (Figure~\ref{logu_calibs}), does offer an explanation for the relatively large scatter between some strong-line ratios and O/H (Figure~\ref{oh_calibs}). Even though our analysis suggests that $\langle z\rangle = 2.3$ KBSS galaxies follow the same N/O-O/H relation as observed in the local universe and that N/O can be accurately and precisely determined for high-$z$ galaxies using common strong-line diagnostics (Figure~\ref{no_calibs}), measuring O/H robustly will remain challenging if it is not also strongly correlated with $U$ (as appears to be the case), especially given the large scatter observed in the N/O-O/H relation.

In addition to the issues of selection bias and model construction we have already highlighted, there may be a physical reason to expect the behavior of $U$ and O/H in high-$z$ galaxies to differ from galaxies and \ion{H}{2} regions in the local universe. \citet{dopita2006} assert that the underlying astrophysical cause for an observed $U$-O/H anti-correlation is the effect of increasing metallicity on the opacity of stellar atmospheres, which will (1) increase the number of ionizing photons absorbed by the stellar wind and (2) more efficiently convert luminous energy flux to mechanical energy needed to launch the wind. However, O/H does not trace enrichment in Fe (which is the primary source of the opacity in stellar atmospheres and is responsible for driving stellar winds) at high redshift in the same manner as in the majority of local galaxies, so the absence of a strong anti-correlation between $U$ and O/H is not especially surprising, if the expected correlation is in fact between $U$ and $Z_{\ast}$.

Figure~\ref{logu_vs_zstar_plot} shows the correlation between $U$ and $Z_{\ast}$, as observed in the KBSS sample. After accounting for measurement errors, we find a 93\% probability of an anti-correlation, with a most-likely linear correlation coefficient of $\rho=-0.20$ signifying a moderately weak relationship. The corresponding best-fit linear relation is shown in cyan and has the form:
\begin{equation}
\log(U) = -3.0-0.27\times \log(Z_{\ast}/Z_{\sun}).
\end{equation}
Our understanding of the intrinsic correlations (or lack thereof) between $U$ and other parameters will benefit from expanding high-$z$ samples to include a larger range in O/H and $Z_{\ast}$, if present, but it seems likely that the paradigm based on our understanding of the local universe is insufficient to fully explain the phenomena observed at high redshift. Fortunately, with samples like KBSS, it is now possible to study the high-$z$ universe independently of calibrations based on low-$z$ galaxy populations.

\subsection{O/Fe in High-$z$ Galaxies}

\begin{figure}
\centering
\includegraphics{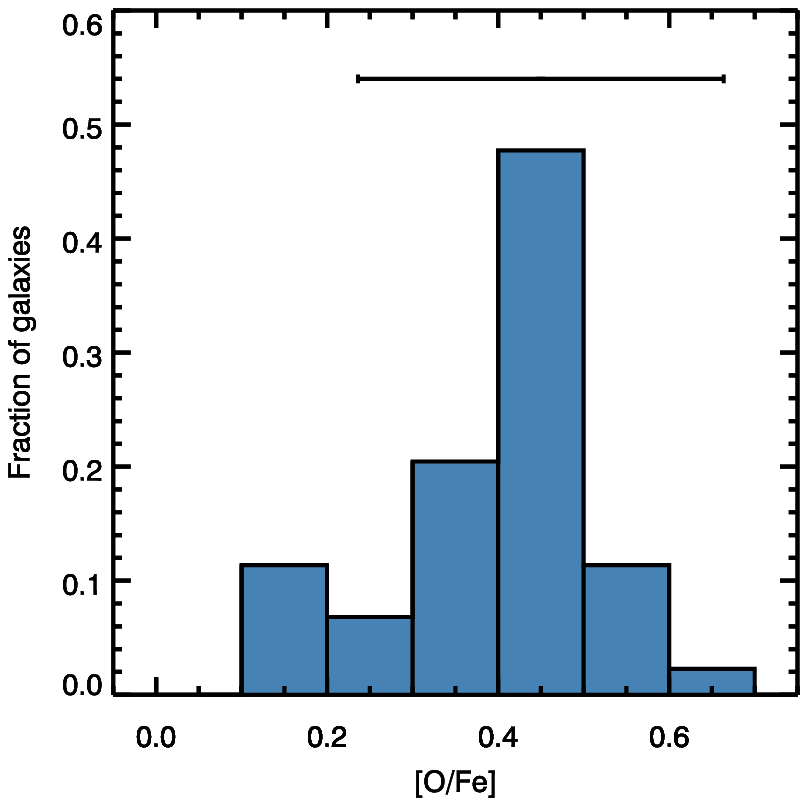}
\caption[Distribution of {[O/Fe]} for $z\simeq2-2.7$ KBSS galaxies]{The distribution of [O/Fe] for the 44 KBSS galaxies with bound posteriors in $\log(Z_{\textrm{neb}}/Z_{\odot})-\log(Z_{\ast}/Z_{\odot})$, including galaxies with limits on N/O. For these galaxies the median [O/Fe]~$=0.42$, which is somewhat lower than the value inferred for a composite spectrum of KBSS galaxies in \citetalias{steidel2016}. Still, these galaxies demonstrate that a significant portion of high-$z$ galaxies exhibit abundance patterns that are indicative of enrichment primarily by CCSNe, modulo uncertainties in supernova yields.}
\label{zratio_hist}
\end{figure}

Just as we translated $\log(Z_{\textrm{neb}}/Z_{\odot})$ into $12+\log(\textrm{O/H})$, we may also directly approximate [O/Fe] for KBSS galaxies using the inferred values of stellar and gas-phase metallicity from the model: $[\textrm{O/Fe}]\approx\log(Z_{\textrm{neb}}/Z_{\odot})-\log(Z_{\ast}/Z_{\odot})$. This ratio roughly traces the relative contributions of Type Ia SNe and CCSNe, which contribute the majority of the Fe and O, respectively, to the overall enrichment of the ISM in galaxies. Figure~\ref{zratio_hist} shows the distribution of [O/Fe] for the 44 KBSS galaxies with bound posteriors in $\log(Z_{\textrm{neb}}/Z_{\odot})-\log(Z_{\ast}/Z_{\odot})$, including galaxies with limits on N/O\footnote{This sample is smaller than the sample shown in Figure~\ref{cloudy_hists} because many galaxies with bound posteriors in the four original model parameters have multiple peaks or limits in the combined $\log(Z_{\textrm{neb}}/Z_{\odot})-\log(Z_{\ast}/Z_{\odot})$ parameter.}. The interquartile range in inferred [O/Fe] for this sample is $0.34-0.48$ with a median [O/Fe]~$=0.42$, somewhat lower than the value we inferred in \citetalias{steidel2016} for a composite spectrum of KBSS galaxies ([O/Fe]~$\sim0.6-0.7$) and also lower than the maximal [O/Fe] expected for purely CCSN enrichment \citep[{[O/Fe]}~$\approx0.73$][]{nomoto2006}. Still, the range in [O/Fe] found in the KBSS subsample shown in Figure~\ref{zratio_hist} is significantly super-solar, reinforcing the need for stellar models that explore non-solar abundance patterns, particularly in O and Fe.

\section{Summary and Conclusions}
\label{summary_section}

We have described a photoionization model method that self-consistently accounts for the primary astrophysical drivers of high-$z$ galaxies' nebular spectra, using observables that are commonly available for most samples. Applying this method to 148 $\langle z\rangle = 2.3$ galaxies from KBSS, we find that:
\begin{itemize}
\item The majority of $\langle z\rangle = 2.3$ galaxies are moderately O-rich, with an interquartile range in $12+\log(\textrm{O/H})=8.29-8.56$ and median $12+\log(\textrm{O/H})=8.37$ (top left panel of Figure~\ref{cloudy_hists}). The same galaxies have significantly sub-solar Fe enrichment, with an interquartile range of [Fe/H]~$=[-0.79,-0.53]$ and median [Fe/H]~$=-0.70$ (bottom left panel of Figure~\ref{cloudy_hists}).
\item There are strong correlations between commonly-used line-ratio indices and both $U$---which is the parameter that galaxies' nebular spectra respond \textit{most} sensitively to---and N/O (Figures~\ref{logu_calibs} and \ref{no_calibs}). We have presented new calibrations for these quantities (Equations~\ref{logu_equation} and \ref{n2o2_equation}), which can be used to estimate $U$ and N/O in high-$z$ galaxies or galaxies with similar stellar populations.
\item The nebular spectra of high-$z$ galaxies resemble more closely those of local \ion{H}{2} regions than the integrated-light spectra of galaxies at $z\sim0$ (Figure~\ref{n2s2_vs_n2o2}). This suggests that diffuse ionized gas may play a less important role in determining the observed rest-optical spectra of high-$z$ galaxies than is thought to be the case for local galaxies \citep[e.g.,][]{sanders2017}.
\item Of the strong-line indices used most commonly for estimating O/H, R23 shows the strongest correlation with gas-phase O/H in high-$z$ galaxies (Figure~\ref{oh_calibs}). However, $\sim84$\% of high-$z$ galaxies in the KBSS sample have values of R23 near the ``turnaround'', making R23 a useful diagnostic only for O-rich systems (Equation~\ref{r23_equation}). A calibration based on both R23 and O32 does not perform significantly better than R23 alone, except at $12+\log(\textrm{O/H})\gtrsim8.6$ (Figure~\ref{o32r23_calib}).
\item The physical parameters inferred from galaxies' nebular spectra are often highly correlated with one another (Figure~\ref{param_correlations}, Table~\ref{covar_table}), complicating studies of the intrinsic relationships between the physical conditions in galaxies. It is therefore necessary to account for both measurement errors and covariance between measurement errors in determining the true correlation between parameters, as we have done in Section~\ref{correlation_section}.
\item High-$z$ galaxies span a similar range in N/O at fixed O/H as observed in a sample of local \ion{H}{2} regions (Figure~\ref{no_vs_oh_plot}). For both samples, N/O varies by up to a factor of $2-3$ at a given O/H, particularly near the transition between the primary plateau and secondary N enrichment.
\item We do not recover a strong anti-correlation between $U$ and either O/H or $Z_{\ast}$, contrary to observations of galaxies in the local universe (Figures~\ref{logu_vs_oh_plot} and \ref{logu_vs_zstar_plot}). Practically, the lack of a strong correlation between these quantities makes it more difficult to infer O/H from high-$z$ galaxies' nebular spectra, which are much more sensitive to changes in $U$.
\item The nebular spectra of most high-$z$ galaxies are consistent with super-solar O/Fe (Figure~\ref{zratio_hist}), indicative of ISM that has been enriched primarily by CCSNe. Given mounting observational evidence for substantially non-solar abundance patterns in elements like O and Fe at high redshift, it is important to compute stellar models that better reflect the enrichment patterns in the early universe---or account for this effect in photoionization models by allowing stellar and nebular ``metallicity" to vary independently of one another, rather than simply scaling solar relative abundances.
\end{itemize}

The general conclusion of this work is that the mapping of abundance patterns to observable properties is significantly different in $z\simeq2-2.7$ galaxies than in the samples of $z\sim0$ galaxies which have heretofore driven the development of diagnostics for inferring galaxy physical conditions. In order to fully characterize galaxies that formed during this critical time in the universe's history---and, in turn, provide constraints on theoretical models of galaxy formation and evolution---it is imperative to avoid making assumptions based on galaxies with significantly different enrichment histories. Although it is not realistic that every observation of individual high-$z$ galaxies will allow for an independent analysis as outlined in this paper, diagnostics based on galaxies with similar stellar populations and star-formation histories should be preferred over locally-calibrated methods. To this end, we have provided guidance for inferring $U$, N/O, and O/H in typical $z\simeq2-2.7$ star-forming galaxies, including re-calibrations of common strong-line diagnostics based on the model results for KBSS galaxies. Finally, we note that the results of applying our photoionization model method also have implications for characterizing other galaxy scaling relations at high redshift, including the mass-metallicity relation (MZR). This will be the focus of forthcoming work.

\acknowledgements
The data presented in this paper were obtained at the W.M. Keck Observatory, which is operated as a scientific partnership among the California Institute of Technology, the University of California, and the National Aeronautics and Space Administration. The Observatory was made possible by the generous financial support of the W.M. Keck Foundation. This work has been supported in part by a US National Science Foundation (NSF) Graduate Research Fellowship (ALS), by the NSF through grants AST-0908805 and AST-1313472 (CCS and ALS). Finally, the authors wish to recognize and acknowledge the significant cultural role and reverence that the summit of Maunakea has within the indigenous Hawaiian community.  We are privileged to have the opportunity to conduct observations from this mountain.

\bibliography{allrefs}

\end{document}